# Optimal Contrast Greyscale Visual Cryptography Schemes with Reversing

Dao-Shun Wang[1,*], Tao Song[1], Lin Dong[1], Ching-Nung Yang[2]

[1]Department of Computer Science and Technology, Tsinghua University, Beijing, 100084, China

[2]Department of Computer Science and Information Engineering, National Dong Hwa University, Taiwan

---

***Abstract:*** Visual cryptography scheme (VCS) is an encryption technique that utilizes human visual system in recovering secret image and it does not require any complex calculation. However, the contrast of the reconstructed image could be quite low. A number of reversing-based VCSs (or *VCSs with reversing*) (RVCS) have been proposed for binary secret images, allowing participants to perform a reversing operation on shares (or shadows). This reversing operation can be easily implemented by current copy machines. Some existing traditional VCS schemes *without* reversing (nRVCS) can be extended to RVCS with the same pixel expansion for binary image, and the RVCS can achieve ideal contrast, significantly higher than that of the corresponding nRVCS. In the application of greyscale VCS, the contrast is much lower than that of the binary cases. Therefore, it is more desirable to improve the contrast in the greyscale image reconstruction. However, when greyscale images are involved, one cannot take advantage of this reversing operation so easily. Many existing greyscale nRVCS cannot be directly extended to RVCS. In this paper, we first give a new greyscale nRVCS with minimum pixel expansion and propose an optimal-contrast greyscale RVCS (GRVCS) by using basis matrices of perfect black nRVCS. Also, we propose an optimal GRVCS even though the basis matrices are not perfect black. Finally, we design an optimal-contrast GRVCS with minimum number of shares held by each participant. The proposed schemes can satisfy different user requirement, previous RVCSs for binary images can be viewed as special cases in the schemes proposed here.

---

[*] Corresponding author. Tel.:+86-10-62782930.
*E-mail address:* daoshun@mail.tsinghua.edu.cn (Daoshun Wang).

# 1. Introduction

A $(k, n)$-visual cryptography scheme (VCS)[1] for black and white image has been proposed to encode a secret image into $n$ "shadow" ("share") images to be distributed to $n$ participants. The secret can be visually reconstructed only when $k$ or more shares are available. No information will be revealed with any $k$ – 1 or fewer shares. The reconstruction process adopts the properties of human visual system without any cryptographic knowledge or operation. In VCS, each secret pixel is subdivided into $m$ subpixels. The value $m$ is named as *pixel expansion*. Based on the definition of [1], Verheul and Van Tilborg [2] gave a more general definition. Suppose that the reconstructed white (resp. black) secret pixel contains $h$ (resp. $l$) white subpixels, where the value of $h$ and $l$ are whiteness of the white and black secret pixels, and $m > h > l \geq 0$. While $l = 0$, i.e. the black pixel can be perfectly reconstructed as $m$ black subpixels, and $h = m$, i.e. the white pixel can be perfectly recovered to white region, such binary VCS has ideal contrast. Blundo et al. [3] introduced how to construct a perfect black $(k, n)$-VCS (PBVCS), which the reconstructed white pixel is not perfectly white region because $m > h$. Blundo et al. [4] gave an estimate of the value of the pixel expansion $m$ of a black and white $(k, n)$-VCS.

To achieve the perfect blackness and the perfect whiteness simultaneously, some researchers consider a totally different approach to improve the quality (contrast) of the recovered image. Viet and Kurosawa [5] noted the phenomenon that most copy machines nowadays have this fundamental function , which can change a black image into white one and vice versa , and then adopted this Boolean Not operation (called reversing) to construct a PBVCS for binary image. In Viet and Kurosawa scheme, the almost ideal contrast of recovered secret image can be obtained for a large number of runs $r$. Cimato et al. [6] presented two elegant construction methods to improve the contrast and pixel expansion of Viet and Kurosawa scheme. To reduce the stacking and revering operations and minimize the number of shadows held by each participant. Hu and Tzeng [7] proposed a novel scheme to construct two ideal contrast VCSs with less reversing and stacking operations in only two runs. In $(k, n)$-Hu and Tzeng's schemes, each participant stores only two shadows (shares), the pixel expansion $2^{k-1}\binom{n}{k}$ is smaller than that of previous deterministic $(k, n)$-VCS schemes [4], when $k \geq \frac{n}{4}$, $k \geq 4$.

Yang et al. [8-9] overcame the weakness of reversing-based perfect VCSs and first introduced a reversing-based scheme for not-perfect black VCS (nPBVCS), this approach uses Boolean XOR operation for decoding. For the convenience of our future discussion, we use “RVCS” to denote this “reversing-based VCS” [8-9] , i.e., “VCS with reversing”. As we know, the XOR operation ‘$\otimes$’ can be reduced as $A \otimes B = \overline{(A \oplus \overline{B})} \oplus \overline{(\overline{A} \oplus B)}$ and implemented by four Not operations and three OR operations ($\oplus$), thus the XOR operation on shares can also be done by a copy machine. Many VCSs with reversing (RVCSs) are accordingly proposed in the literatures [10-12]. Tan [10] gave a (2, 2)-RVCS mixed on XOR operation and OR operation at first, and then proposed a (*k*, *n*) secret sharing scheme based on binary linear error-correcting code. Zhang et al.[11] proposed a novel ideal contrast RVCS based on probabilistic VCS. Fang et al.[12] presented a novel multi-secret RVCS.

A RVCS is called fully compatible [5-9] if the participants can still recover the secret image without a copy machine in the reconstruction phases. Valid factors to be considered for designing RVCSs include compatibility, complexity of reconstructed secret image, number of shares held by each participant, number of runs to achieve perfect contrast, contrast, pixel expansion, and variant aspect ratio. Those factors of typical schemes for binary image are shown in Table A-1 of Appendix A. In this Table A-1, we do not include Cimato et al’s second scheme because it does not provide the compatibility. From Table A-1, Cimato et al’s scheme is optimal with contrast and no variant aspect ratio. Hu and Tzeng’s scheme is optimal with minimum number of shares held by each participant and low complexity for reconstruction. Yang et al.’s B scheme is optimal for nPBVCS.

Directly based on binary schemes, VCSs for grey-scale images (called GVCS) [13-14] with optimal pixel expansion $m_g = (g-1) \cdot m$ are proposed, and the contrast between two neighboring grey levels is $\frac{1}{(g-1)m}$. The almost optimal pixel expansion can be achieved in VCSs for binary images and grey-scale images. For example, when a (3, 3)-GVSS scheme proposed in [13-14] is used to code an image with 256 grey-levels, the contrast is as small as

1/1024.

In this case, it is straightforward to construct a greyscale VCS without reversing (nRGVCS) to improve its contrast. Without reversing, binary nRVCS can be directly generalized to construct greyscale GVCS. With reversing, however, we cannot directly extend the existing typical binary RVCS to construct greyscale schemes. This point is illustrated in Section 3, 4, and 5 of this paper.

In section 2, we briefly review binary VCS and grayscale image VCS, and obtain the condition of ideal contrast in greyscale VCS. In section 3, we analyze the reasons why existing typical Cimato et al.' binary RVCS [6] cannot be extended to greyscale VCS, and construct a new grayscale nRVCS. Then, we propose an optimal-contrast greyscale RVCS (GRVCS) by using basis matrices of PBVCS. In section 4, we propose an optimal greyscale reversing-based VCS even though the basis matrices are not perfect black. In section 5, we design an optimal-contrast GRVCS with minimum number of shares held by each participant. Comparisons and discussions are given in section 6 and the conclusions are given in section 7.

## 2. Background, Preliminaries, Basic results

This section briefly reviews traditional visual cryptography scheme (VCS) [1-2] and Blundo et al. and Mucke's greyscale visual cryptography scheme(GVCS) [13-14].Some basic notations are defined when they first appear in the text and a list of important notations is given in Table A-2 and A-3 of Appendix A.

### *2.1 Traditional binary (k, n)-VCS*

In a binary VCS, the secret image consists of a collection of black-and-white pixels and each pixel is subdivided into a collection of $m$ black-and-white subpixels in each of the $n$ shares. The collection of subpixels can be represented by an $n \times m$ Boolean matrix $S$=[$s_{ij}$], where the element $s_{ij}$ represents the *j-th* subpixel in the *i-th* share. A white pixel is represented as a 0, and a black pixel is represented as a 1. On a transparency, white subpixels allow light to pass through while black subpixels stop light. One has that $s_{ij}$ = 1 if and only if the *j-th* pixel in the *i-th* share is black. Stacking shares $i_1, \ldots, i_r$ together, the grey-level of each pixel ($m$ subpixels) of the combined share is proportional to the Hamming weight (the number of 1's in the vector $V$) $H(V)$ of the OR-ed ("OR" operation) $m$-vector $V = OR(i_1, \cdots, i_r)$ where $i_1$, …, $i_r$ are the rows of $S$ associated with the shares we stack. Verheul and Van Tilborg [2] extended the definition of Naor and Shamir's scheme[1].

The formal definition of binary VCS is given below.

***Definition 2.1[2]:*** A solution to the *k* out of *n* visual cryptography scheme consists of two collections of $n \times m$ Boolean matrices $B_0$ and $B_1$. To share a white (resp. black) pixel, the dealer randomly chooses one of the matrices in $B_0$ (resp. $B_1$). The chosen matrix defines the color of the $m$ subpixels in each one of the *n* transparencies. The solution is considered valid if the following three conditions are met.

1. For any $S$ in $B_0$, the OR vector $V^0$ of any *k* of the *n* rows satisfies $H(V^0) \le m - h$, $h \in Z^+$.
2. For any $S$ in $B_1$, the OR vector $V^1$ of any *k* of the *n* rows satisfies $H(V^1) \ge m - l$, $l \in Z^+$, $l < h \le m$.
3. For any subset $\{i_1, \ldots, i_q\}$ of $\{1, \ldots, n\}$ with $q < k$, the two collections of $q \times m$ matrices $D_t$ for $t \in \{0, 1\}$ obtained by restricting each $n \times m$ matrix in $B_t$ (where $t = 0, 1$) to rows $i_1, \ldots, i_q$ are indistinguishable in the sense that they contain the same matrices with the same frequencies.

The first two conditions are called "contrast" and the third condition is called "security". In this definition, the parameter *m* is called *pixel expansion*, which refers to the number of subpixels representing a pixel in the secret image. The *contrast* $\alpha = (H(V^1) - H(V^0))/m = (h - l)/m$, also called *relative difference*, refers to the difference in weight between combined shares that come from a white pixel and a black pixel in the secret image. When $\alpha = 1$, the contrast is said to be ideal.

From Definition 2.1, a binary (*k*, *n*)-VCS can be realized by the two Boolean matrices $B_0$ and $B_1$. The collection $C_0$ (resp. $C_1$) can be obtained by permuting the columns of the corresponding Boolean matrix $B_0$ (resp. $B_1$) in all possible ways. $B_0$ and $B_1$ are called basis matrices, and hence each collection has $m!$ matrices.

Let $OR(B_i|t)$ denote the "OR"-ed $t$ rows in $B_i$ $i$ =0, 1, and $H(.)$ be the Hamming weight function. We can re-write Definition 2.1 as follows.

(D-1) $H(OR(B_0|t)) \le m - h$ and $H(OR(B_1|t)) \ge m - l$ for $t \ge k$, where $0 \le l < h \le m$.

(D-2) $H(OR(B_0|t)) = H(OR(B_1|t))$ for $t \le k - 1$.

The next example lets us to understand the definition 2.1 above.

***Example 2.1:*** Suppose in a (2, 3)-VCS, each pixel in a secret image is encoded into a collection of 3 black and white subpixels in each share of the 3 shares. The encoding matrices can be represented by two $3 \times 3$ 0/1 matrices (from [23]):

$$B_0 = \begin{bmatrix} 0 & 1 & 1 \\ 0 & 1 & 1 \\ 0 & 1 & 1 \end{bmatrix}, \quad B_1 = \begin{bmatrix} 0 & 1 & 1 \\ 1 & 0 & 1 \\ 1 & 1 & 0 \end{bmatrix}$$

where 0 ( resp.1 ) denotes a white subpixel ( reps. a black subpixel ).

Let $C_0$ ($C_1$ ) be the collection of all matrices obtained by permuting all columns of $B_0$ ($B_1$ ), namely

$$C_0 = \left\{ \begin{bmatrix} 0 & 1 & 1 \\ 0 & 1 & 1 \\ 0 & 1 & 1 \end{bmatrix}, \begin{bmatrix} 1 & 0 & 1 \\ 1 & 0 & 1 \\ 1 & 0 & 1 \end{bmatrix}, \begin{bmatrix} 1 & 1 & 0 \\ 1 & 1 & 0 \\ 1 & 1 & 0 \end{bmatrix} \right\},$$

$$C_1 = \left\{ \begin{bmatrix} 0 & 1 & 1 \\ 1 & 0 & 1 \\ 1 & 1 & 0 \end{bmatrix}, \begin{bmatrix} 0 & 1 & 1 \\ 1 & 1 & 0 \\ 1 & 0 & 1 \end{bmatrix}, \begin{bmatrix} 1 & 0 & 1 \\ 0 & 1 & 1 \\ 1 & 1 & 0 \end{bmatrix}, \begin{bmatrix} 1 & 1 & 0 \\ 0 & 1 & 1 \\ 1 & 0 & 1 \end{bmatrix}, \begin{bmatrix} 1 & 1 & 0 \\ 1 & 0 & 1 \\ 0 & 1 & 1 \end{bmatrix}, \begin{bmatrix} 1 & 0 & 1 \\ 1 & 1 & 0 \\ 0 & 1 & 1 \end{bmatrix} \right\}.$$

Obviously, $H(OR(B_0|2)) = 2$ and $H(OR(B_1|2)) = 3$ satisfy Condition (D-1), and $H(OR(B_0|1)) = H(OR(B_1|1) = 2$ satisfies Condition (D-2). The contrast is $\alpha = (h-l)/m = 1/3$.

### *2.2 Greyscale visual cryptography scheme*

In the greyscale model, the original image has a greysacle palette with $g$ distinct grey levels, where $g \geq 2$. A primary color will have an intensity range between 0 and 1, with 0 representing white and 1 represents black. Directly based on binary VCSs, Muecke [13] and Blundo et al. [14] independently presented a general approach to implement grey-levels VCS.

***Definition 2.2[13-14]:*** A solution to the *k* out of *n* visual secret sharing scheme consists of a family of $g\,(g \geq 2)$ collections of $n \times m_g$ Boolean matrices $\{G^0, \cdots, G^{g-1}\}$, where $G^q$ is the collection for grey level $i_q$ for $0 \leq q \leq g-1$. To share a pixel with a grey level of $i_q$, the dealer randomly chooses a $n \times m_g$ matrix from the matrices $G^q$ to define the color of the $m_g$ subpixels in each one of the $n$ transparencies. If there exist a set of contrast $\{\alpha^{(1,0)}, \cdots, \alpha^{(g-1,g-2)}\}$ ,where $\alpha^{(i+1,i)}$ is contrast between $i$ -grey level and $i+1$ grey level, $i = 0, \cdots, g-2$ and sets of threshold $\{d_1, \cdots, d_{g-1}\}$.The solution is considered valid if the following three conditions are met.

1. For any $S^q \in G^q$, the Hamming distance between the OR $m_g$-vector $V^q$ of any *k* of the $n$ rows in $S^q$ satisfies $H(V^q) \le d_i - \alpha^{(i+1,i)} \times m_g$
2. For any $S^{q+1} \in G^{q+1}$, the Hamming distance between the the OR $m_g$-vector $V^{q+1}$ of any *k* of the $n$ rows in $S^{q+1}$ satisfies $H(V^{q+1}) \ge d_i$
3. For any subset $\{i_1, \cdots, i_t\}$ of $\{1, \cdots, n\}$ with $t < k$, the collections of $t \times m_g$ matrices obtained by restricting each $n \times m_g$ matrix in $G^q$ to rows $i_1, \cdots, i_t$ are indistinguishable in the sense that they contain the same matrices with the same frequencies for $0 \le q \le g-1$.

The first two conditions ensure that contrast is maintained between grey levels. It states that two neighboring entries in the greyscale palette must have a relative contrast difference of at least $\alpha^{(i+1,i)} > 0$, $i = 0, \cdots, g-2$.

The third condition ensures the security of the scheme. It states that if less than *k* shares are stacked together, we will not be able to determine which collection the matrix was selected from. Therefore, we will not be able to determine the color of the original pixel.

Let $OR(G^q|t)$ denote the "OR"-ed $t$ rows in $G^q (1 \le q \le g)$.We can re-write Definition 2.2 as follows.

(D-3) $H(OR(G^q|t)) \le d_i - \alpha^{(i+1,i)} \times m_g$ and $H(OR(G^{q+1}|t)) \ge d_i$ for $t \ge k$

(D-4) $H(OR(G^{q+1}|t)) = H(OR(G^q|t))$ for $t \le k-1$.

Let $A = [a_{ij}]_{n \times m}$ and $B = [b_{ij}]_{n \times m}$ be two basis matrices with $n \times m$ size. Let the symbol "$\circ$" denote the concatenation operation, which describes the relation of the combination of two basis Boolean matrices, i.e. $A \circ B = [a_{ij}]_{n\times m} \circ [b_{ij}]_{n\times m} = [a_{ij}\ b_{ij}]_{n\times 2m}$. It is easy to see that the order of the two basis matrices does not affect the combination result. Indeed, "$\circ$" is a commutative operation, i.e. $A \circ B = B \circ A$.

We use a binary (2, 3)-VCS to construct a $(2,3)$-GVCS with three grey levels.

***Example 2.2(continuation of Example 1) [13-14] :***

$B_0$ and $B_1$ are basis matrices of a binary (2, 3)-VCS.

$$B_0 = \begin{bmatrix} 1 & 1 & 0 \\ 1 & 1 & 0 \\ 1 & 1 & 0 \end{bmatrix},\ B_1 = \begin{bmatrix} 0 & 1 & 1 \\ 1 & 0 & 1 \\ 1 & 1 & 0 \end{bmatrix}.$$

We have

$$G^0 = B_0 \circ B_0 = \begin{bmatrix} 0 & 1 & 1 \\ 0 & 1 & 1 \\ 0 & 1 & 1 \end{bmatrix} \circ \begin{bmatrix} 0 & 1 & 1 \\ 0 & 1 & 1 \\ 0 & 1 & 1 \end{bmatrix} = \begin{bmatrix} 0 & 1 & 1 & 0 & 1 & 1 \\ 0 & 1 & 1 & 0 & 1 & 1 \\ 0 & 1 & 1 & 0 & 1 & 1 \end{bmatrix}, G^1 = B_0 \circ B_1 = \begin{bmatrix} 0 & 1 & 1 \\ 0 & 1 & 1 \\ 0 & 1 & 1 \end{bmatrix} \circ \begin{bmatrix} 0 & 1 & 1 \\ 1 & 0 & 1 \\ 1 & 1 & 0 \end{bmatrix} = \begin{bmatrix} 0 & 1 & 1 & 0 & 1 & 1 \\ 0 & 1 & 1 & 1 & 0 & 1 \\ 0 & 1 & 1 & 1 & 1 & 0 \end{bmatrix},$$

$$G^2 = B_1 \circ B_1 = \begin{bmatrix} 0 & 1 & 1 \\ 1 & 0 & 1 \\ 1 & 1 & 0 \end{bmatrix} \circ \begin{bmatrix} 0 & 1 & 1 \\ 1 & 0 & 1 \\ 1 & 1 & 0 \end{bmatrix} = \begin{bmatrix} 0 & 1 & 1 & 0 & 1 & 1 \\ 1 & 0 & 1 & 1 & 0 & 1 \\ 1 & 1 & 0 & 1 & 1 & 0 \end{bmatrix}$$

Since $H(OR(G^0|2) = 4$, $H(OR(G^1|2) = 5$, and $H(OR(G^2|2) = 6$, thus it satisfies Condition (D-3). In addition, $H(OR(G^2|1) = H(OR(G^1|1) = H(OR(G^0|1) = 4$ satisfies Condition (D-4). The contrast is

$$\alpha^{(1,0)} = \frac{H(C^{(1)}|2) - H(C^{(0)}|2)}{m_g} = \frac{5-4}{6} = \frac{1}{6}, \alpha^{(2,1)} = \frac{H(C^{(2)}|2) - H(C^{(1)}|2)}{m_g} = 1 - \frac{5}{6} = \frac{1}{6}.$$

### *2.3 Some basic results for greyscale (k, n)-VCS*

The following Theorem 1 is immediately gotten from Theorem 3.2 of Blundo et al.'s result [14].

***Theorem 2.1[14] :*** Let $\alpha^*$ be the maximum contrast (or relative difference) of a $(k,n)$-GVCS with $g$ ($g \geq 1$) grey levels. There exists a $(k,n)$-GVCS with contrast $\alpha^{(1,0)}, \cdots, \alpha^{(g-1,g-2)}$ if and only if $\sum_{i=0}^{g-2} \alpha^{(i+1,i)} \leq a^*$.

The next Lemma 2.1 is an immediate consequence of Theorem 2.1 above.

***Lemma 2.1[14]:*** In $(k,n)$-GVCS, with contrast $\alpha^{(1,0)}, \cdots, \alpha^{(g-1,g-2)}$, it holds that

$\min\{\alpha^{(1,0)},\cdots,\alpha^{(g-1,g-2)}\}\leq\frac{1}{(g-1)\cdot m_g}$ and $m_g \geq (g-1)\cdot m$, where $m$ is pixel expansion of a binary $(k,n)$-VCS.

Using Lemma 2.1, we obtain the following corollary 2.1.

***Corollary 2.1[14]:*** In $(k,k)$-GVCS with contrast $\alpha^{(1,0)},\cdots,\alpha^{(g-1,g-2)}$, the $\min\{\alpha^{(1,0)},\cdots,\alpha^{(g-1,g-2)}\}\leq\frac{1}{(g-1)\cdot m_g}$ and $m_g \geq (g-1)\cdot 2^{k-1}$, where $m$ is minimum pixel expansion of a binary $(k,k)$-VCS [1].

As we know, in a binary $(k,n)$-VCS, the idea contrast $\alpha=1$. Using Theorem 2.1 above, we obtain the following Theorem 2.

***Theorem 2.2:*** In a greyscale $(k,n)$-VCS with contrast $\alpha^{(1,0)},\cdots,\alpha^{(g-1,g-2)}$, it holds that optimal contrast $\alpha^{(1,0)}=\cdots=\alpha^{(g-1,g-2)}=\frac{1}{(g-1)}$.

**Proof:** When $\alpha^{(1,0)}=\cdots=\alpha^{(g-1,g-2)}$.By using the result of theorem 2.1 above, $\sum_{i=0}^{g-2}\alpha^{(i+1,i)}\leq a^*$, when $\alpha^*\equiv 1$,then we obtain the ideal contrast $\alpha^{(1,0)}=\cdots=\alpha^{(g-1,g-2)}=\frac{1}{g-1}$.□

***Theorem 2.3 ([13-14]):*** In $(k,n)$-VCS with $g$ grey levels, the pixel expansion $m_g$ and the contrast $\alpha^{(i+1,i)}$ between grey levels are

$$m_g = (g-1)\cdot m,$$

$$\alpha^{(i+1,i)}=\frac{\alpha}{g-1},\quad i=0,\cdots,g-2.$$

Notice that the two parameteres $m$ and $\alpha$ in the theorem above are pixel expansion and contrast of a binary $(k,n)$-VCS.

# 3. The proposed greyscale reversing-based VCS by using PBVCS

In this section we present more detailed analysis to Cimato et al.’s scheme [6], which cannot be directly extended to grayscale RVCS. Based on some of the ideas from Cimato et al.’s binary RVCS, we first device a permutation operation for basis matrices, then construct a new grayscale scheme without reversing (nRGVCS), and finally propose a corresponding reversing-based grayscale VCS (RGVCS) with optimal-contrast.

### *3.1 The analysis of directly extending RVCS to RGVCS*

Muecke [13] and Blundo et al. [14] independently presented a general approach to implement grayscale VCS based on binary VCSs. A natural extension for a binary VCS with reversing is to a grayscale image whose pixels have $g$ grey levels ranging from 0 (representing a white pixel) to $g-1$ (representing a black pixel).

Cimato et al.'s perfect black VCS (PBRVCS) [6] can only be used for binary images because it uses a single bit to represent each pixel. In **Example B-2 of** Appendix B, we directly use Cimato et al.'s binary PBRVCS to perform three grey levels (2, 3)-GVCS with reversing. From the experimental result we can see that the secret image cannot be reconstructed correctly. Greyscale images with more than two grey levels can not directly benefit from Cimato et al.'s method.

For images with three or more grey levels, each pixel must correspond to a string of multiple bits. How many bits should be used to represent $g$ different grey levels? We now give a result, which looks simple but is a very useful conclusion to create a more general (*k*, *n*)-VCS and reversing based VCS for grayscale scheme.

***Lemma 3.1:*** In a RVCS, a pixel with $g$ different grey-levels needs at least a binary string of $g-1$ bits to represent their value.

***Proof:*** A binary string of $g-1$ bits have $2^{g-1}$ cases (states), namely $2^{g-1}=\binom{g-1}{0}+\binom{g-1}{1}+\ldots+\binom{g-1}{g-1}$. As we know, a binary string of $g-1$ bits can be converted into a row vector of $g-1$ components. The corresponding Hamming weights of the binary string are 0, 1, 2, .., $g-1$, respectively. Therefore, a binary string of $g-1$ bits can have as many as $g$ different Hamming weights, starting from all $g-1$ zeros to all ones. If each distinct Hamming weight can be used to represent a unique grey-level, as in the case of transparency overlay, at least $g-1$ bits must be involved in coding a grayscale image with $g$ different grey-levels.

For example, when g=4, a binary string of 3 bits has 8 different combinations of 0s and 1s, including one combination (“000”) with Hamming weight 0, one combination (“111”) with Hamming weight 3, three with Hamming weight 1 (“001”, “010”, and “100”), and three with Hamming weight 2 (“110”, “101”, “011”).□

While $m$ is minimum value for a binary (*k*, *n*)-nRVCS, then the lemma 3.1 also shows Blundo et al.’ grayscale nRVCS [14] has minimum pixel expansion.

### *3.2 The relationship between contrast and column permutation method*

Using the result of Lemma 3.1, we now apply 2 bits to extend Cimato et al.’s idea as follows.

***Example 3.1*(continuation of Example 2.2):** In a three grey levels (2, 3)-GVCS with $m_g = 6$ under different permutation methods. The basis matrices are:

$$G^0 = B_0 \circ B_0 = \begin{bmatrix} 0 & 1 & 1 \\ 0 & 1 & 1 \\ 0 & 1 & 1 \end{bmatrix} \circ \begin{bmatrix} 0 & 1 & 1 \\ 0 & 1 & 1 \\ 0 & 1 & 1 \end{bmatrix} = \left[ \begin{array}{ccc|ccc} 0 & 1 & 1 & 0 & 1 & 1 \\ 0 & 1 & 1 & 0 & 1 & 1 \\ \underbrace{0 \quad 1 \quad 1}_{Component1} & & & \underbrace{0 \quad 1 \quad 1}_{Component2} & & \end{array} \right] \begin{array}{l} \to 1 \\ \to 2 \\ \to 3 \end{array}$$

$$G^1 = B_0 \circ B_1 = \begin{bmatrix} 0 & 1 & 1 \\ 0 & 1 & 1 \\ 0 & 1 & 1 \end{bmatrix} \circ \begin{bmatrix} 0 & 1 & 1 \\ 1 & 0 & 1 \\ 1 & 1 & 0 \end{bmatrix} = \left[ \begin{array}{ccc|ccc} 0 & 1 & 1 & 0 & 1 & 1 \\ 0 & 1 & 1 & 1 & 0 & 1 \\ \underbrace{0 \quad 1 \quad 1}_{Component1} & & & \underbrace{1 \quad 1 \quad 0}_{Component2} & & \end{array} \right] \begin{array}{l} \to 1 \\ \to 2 \\ \to 3 \end{array}$$

$$G^2 = B_1 \circ B_1 = \begin{bmatrix} 0 & 1 & 1 \\ 1 & 0 & 1 \\ 1 & 1 & 0 \end{bmatrix} \circ \begin{bmatrix} 0 & 1 & 1 \\ 1 & 0 & 1 \\ 1 & 1 & 0 \end{bmatrix} = \left[ \begin{array}{ccc|ccc} 0 & 1 & 1 & 0 & 1 & 1 \\ 1 & 0 & 1 & 1 & 0 & 1 \\ \underbrace{1 \quad 1 \quad 0}_{Component1} & & & \underbrace{1 \quad 1 \quad 0}_{Component2} & & \end{array} \right] \begin{array}{l} \to 1 \\ \to 2 \\ \to 3 \end{array}$$

**Permutation Method I:**

For every pixel, we randomly choose matrices, which are gotten by doing totally random column permutation to the basis matrices, to distribute the pixel in each share. We denote this totally random column permutation as Permutation Method I. Since there are three grey levels, we use two bits to represent the pixel in each share (see Lemma 3.1). For different chosen matrices, we give the following two situations.

**Situation 1**: suppose the chosen matrices are

$$S^0 = \begin{bmatrix} 0 & 1 & 0 & 1 & 1 & 1 \\ 0 & 1 & 0 & 1 & 1 & 1 \\ 0 & 1 & 0 & 1 & 1 & 1 \end{bmatrix},\ S^1 = \begin{bmatrix} 1 & 1 & 1 & 1 & 0 & 0 \\ 1 & 0 & 1 & 1 & 0 & 1 \\ 1 & 1 & 1 & 0 & 0 & 1 \end{bmatrix},\ S^2 = \begin{bmatrix} 1 & 1 & 1 & 1 & 0 & 0 \\ 0 & 0 & 1 & 1 & 1 & 1 \\ 1 & 1 & 0 & 0 & 1 & 1 \end{bmatrix}$$

Table 3.1 The distribution phase under Situation 1

| Grey levels | Chosen matrices | 1st run | 2nd run | 3rd run |
|---|---|---|---|---|
| 1 | $\begin{bmatrix} 0 & 1 & 0 & 1 & 1 & 1 \\ 0 & 1 & 0 & 1 & 1 & 1 \\ 0 & 1 & 0 & 1 & 1 & 1 \end{bmatrix}$ | $t_{1,1}$ =01 | $t_{1,2}$ =01 | $t_{1,3}$ =11 |
| | | $t_{2,1}$ =01 | $t_{2,2}$ =01 | $t_{2,3}$ =11 |
| | | $t_{3,1}$ =01 | $t_{3,2}$ =01 | $t_{3,3}$ =11 |
| 2 | $\begin{bmatrix} 1 & 1 & 1 & 1 & 0 & 0 \\ 1 & 0 & 1 & 1 & 0 & 1 \\ 1 & 1 & 1 & 0 & 0 & 1 \end{bmatrix}$ | $t_{1,1}$ =11 | $t_{1,2}$ =11 | $t_{1,3}$ =00 |
| | | $t_{2,1}$ =10 | $t_{2,2}$ =11 | $t_{2,3}$ =01 |
| | | $t_{3,1}$ =11 | $t_{3,2}$ =10 | $t_{3,3}$ =01 |
| 3 | $\begin{bmatrix} 1 & 1 & 1 & 1 & 0 & 0 \\ 0 & 0 & 1 & 1 & 1 & 1 \\ 1 & 1 & 0 & 0 & 1 & 1 \end{bmatrix}$ | $t_{1,1}$ =11 | $t_{1,2}$ =11 | $t_{1,3}$ =00 |
| | | $t_{2,1}$ =00 | $t_{2,2}$ =11 | $t_{2,3}$ =11 |
| | | $t_{3,1}$ =11 | $t_{3,2}$ =00 | $t_{3,3}$ =11 |

Table 3.2 The reconstruction phase by participant 1 and participant 2 under Situation 1

| Grey levels | $T_1 = OR(t_{1,1}, t_{2,1})$ | $T_2 = OR(t_{1,2}, t_{2,2})$ | $T_3 = OR(t_{1,3}, t_{2,3})$ | $\tilde{P} = \overline{(OR(\overline{T_1}, \overline{T_2}, \overline{T_3}))}$ |
|---|---|---|---|---|
| 1 | 01 | 01 | 11 | 01 |
| 2 | 11 | 11 | 01 | 01 |
| 3 | 11 | 11 | 11 | 11 |

**Situation 2**: suppose the chosen matrices are

$$S^0 = \begin{bmatrix} 0 & 0 & 1 & 1 & 1 & 1 \\ 0 & 0 & 1 & 1 & 1 & 1 \\ 0 & 0 & 1 & 1 & 1 & 1 \end{bmatrix},\ S^1 = \begin{bmatrix} 1 & 1 & 0 & 1 & 1 & 0 \\ 1 & 1 & 0 & 0 & 1 & 1 \\ 1 & 1 & 0 & 1 & 0 & 1 \end{bmatrix},\ S^2 = \begin{bmatrix} 1 & 1 & 0 & 1 & 1 & 0 \\ 0 & 1 & 1 & 0 & 1 & 1 \\ 1 & 0 & 1 & 1 & 0 & 1 \end{bmatrix}$$

Table 3.3 The distribution phase under Situation 2

| Grey levels | Chosen matrices | 1st run | 2nd run | 3rd run |
|---|---|---|---|---|
| 1 | $\begin{bmatrix} 0 & 0 & 1 & 1 & 1 & 1 \\ 0 & 0 & 1 & 1 & 1 & 1 \\ 0 & 0 & 1 & 1 & 1 & 1 \end{bmatrix}$ | $t_{1,1}$ =00 | $t_{1,2}$ =11 | $t_{1,3}$ =11 |
| | | $t_{2,1}$ =00 | $t_{2,2}$ =11 | $t_{2,3}$ =11 |
| | | $t_{3,1}$ =00 | $t_{3,2}$ =11 | $t_{3,3}$ =11 |
| 2 | $\begin{bmatrix} 1 & 1 & 0 & 1 & 1 & 0 \\ 1 & 1 & 0 & 0 & 1 & 1 \\ 1 & 1 & 0 & 1 & 0 & 1 \end{bmatrix}$ | $t_{1,1}$ =11 | $t_{1,2}$ =01 | $t_{1,3}$ =10 |
| | | $t_{2,1}$ =11 | $t_{2,2}$ =00 | $t_{2,3}$ =11 |
| | | $t_{3,1}$ =11 | $t_{3,2}$ =01 | $t_{3,3}$ =01 |
| 3 | $\begin{bmatrix} 1 & 1 & 0 & 1 & 1 & 0 \\ 0 & 1 & 1 & 0 & 1 & 1 \\ 1 & 0 & 1 & 1 & 0 & 1 \end{bmatrix}$ | $t_{1,1}$ =11 | $t_{1,2}$ =01 | $t_{1,3}$ =10 |
| | | $t_{2,1}$ =01 | $t_{2,2}$ =10 | $t_{2,3}$ =11 |
| | | $t_{3,1}$ =10 | $t_{3,2}$ =11 | $t_{3,3}$ =01 |

Table 3.4 The reconstruction phase by participant 1 and participant 2 under Situation 2

| Grey levels | $T_1 = OR(t_{1,1}, t_{2,1})$ | $T_2 = OR(t_{1,2}, t_{2,2})$ | $T_3 = OR(t_{1,3}, t_{2,3})$ | $\tilde{P} = \overline{(OR(\overline{T_1}, \overline{T_2}, \overline{T_3}))}$ |
|---|---|---|---|---|
| 1 | 00 | 11 | 11 | 00 |
| 2 | 11 | 01 | 11 | 01 |
| 3 | 11 | 11 | 11 | 11 |

We can see that in Situation 1, both grey level 1 and grey level 2 are reconstructed as "01", thus we cannot distinguish them. While in Situation 2, three different grey levels are reconstructed as different bits. These two situations demonstrate that this scheme is a probabilistic scheme. Actually, grey level 1 will be reconstructed as "01" or "10" with probability 40% and will be reconstructed as "00" with probability 60%. That is to say, grey level 1 cannot be reconstructed correctly.

From the above experiments and analysis, we can see that we cannot get a grayscale scheme with optimal contrast by doing totally random column permutations to the basis matrices.

We consider some special column permutation where only local column exchanges are allowed, and denote this collection as $\overset{\leftrightarrow}{G^q}$. There are two kinds of local column permutation methods.

**Permutation Method II:**

Permutation Method II is that columns of each Component are permutated within that Component, no columns of different Components are exchanged, and all Components go through exactly the same internal column permutation simultaneously (see Example B-3 **of** Appendix B).

**Permutation Method III:**

Permutation Method III is that columns of each Component are permutated within that Component independently; no columns of different Components are exchanged (see Example B-4 **of** Appendix B).

**Example B-5 of** Appendix B, which is a (2, 3)-GVCS with 3 grey levels, demonstrates the difference between "local" and "global" permutation or exchange.

From the experimental result we can see that both methods can reconstruct the secret image correctly, but some secret information leaks out in Permutation Method II. Permutation method III can reconstruct the secret image correctly and each share doesn't leak out any information. The reason is that every component is random permuted separately.

Building on experimental results and analysis above, the existing GVCS cannot be easily extended to RGVCS . Therefore, it is necessary to construct a new GVCS and then extend this GVCS to RGVCS. We now propose a "*within-block-column- permutation*" method (referred to as "WBCP" later) for basis matrices, and then design a new GVCS using WBCP.
We will use nR-WBCP -GVCS to represent our scheme for grayscale $(k,n)$ -GVCS within-block-column- permutation. We will also show that the nR-WBCP -GVCS is different from traditional grayscale nRVCS (nRGVCS) in their permutation methods.

### 3.3 The proposed grayscale $(k,n)$ -GVCS within-block-column- permutation

Let $B_0$ and $B_1$ be the basis matrices for a binary $(k,n)$-VCS scheme with pixel expansion $m$ and contrast $\alpha$. Next we will give a new construction of $(k,n)$-GVCS (nRB-GVCS) with $g$ grey levels.

---

**Construction 3.1:** Construct a new $(k,n)$-GVCS based on a binary $(k,n)$-nRVCS.

**Input:** basis matrices $B_0$ and $B_1$ of the $(k,n)$-nRVCS

**Output:** matrix collection $\overset{\leftrightarrow}{C}^{(q)}$ of the $(k,n)$-nRVCS with $g$ grey levels.

**Construct procedure**:

**Step 1:** Let $\Gamma_{Grey} = \{i_1, \cdots, i_g\}$ be a greyscale palette with $g \ge 2$ grey levels.

**Step 2:** Let $B_t$ be basis matrix of $n \times m$, where $t = 0,1$. Let $\overset{\leftrightarrow}{G}^q$ be a matrix of $n \times ((g-1)\cdot m)$, constructed by concatenating $(g-1)$ $B_t$ matrixes, where $g-q-1$ of them are $B_0$'s, and $q$ of them $B_1$'s . That

$$\overset{\leftrightarrow}{G}^q = \overbrace{\overset{\leftrightarrow}{B}_0 \circ \cdots \circ \overset{\leftrightarrow}{B}_0}^{g-q-1} \circ \overbrace{\overset{\leftrightarrow}{B}_1 \circ \cdots \circ \overset{\leftrightarrow}{B}_1}^{q}$$ ,where $0 \le q \le g-1$.

Step 3: $\overset{\leftrightarrow}{C}^{(q)}$ is obtained by permuting the columns of $\overset{\leftrightarrow}{G}^q$ in such a way: the columns of each "Component", which is $B_0$ or $B_1$, are permutated within that Component, no columns of different Components are exchanged.

---

In Construction 3.1 above, the pixel expansion $m_g = (g-1)m$. The symbol $\leftrightarrow$ represents random column permutation of basis matrices $B_0$ and $B_1$ is only restricted within $B_0$ and $B_1$, and no columns of different basis matrices $B_0$ and $B_1$ are exchanged. In the special case basis matrix for white (i.e. grey level 1) pixel is $\overset{\leftrightarrow}{G}^0 = \overbrace{\overset{\leftrightarrow}{B}_0 \circ \cdots \circ \overset{\leftrightarrow}{B}_0}^{g-1}$ and for black (i.e. grey level $g$ ) pixel is $\overset{\leftrightarrow}{G}^{g-1} = \overbrace{\overset{\leftrightarrow}{B}_1 \circ \cdots \circ \overset{\leftrightarrow}{B}_1}^{g-1}$ .

The collection $\overset{\leftrightarrow}{C}^{(q)}$ is obtained by permuting the columns of $\overset{\leftrightarrow}{G}^q$ . If all possible column permutations are permitted, we end up with $m_g!$ matrices. Here, we consider a specific subset of permutations where the columns of each "Component", which is $B_0$ or $B_1$, are permutated within that Component, and no columns of different Components are exchanged. This restricted column permutation produces $(m!)^{g-1}$ matrices in the collection $\overset{\leftrightarrow}{C}^{(q)}$ .

Let $V^q$ be $t$ row vectors in basis Boolean matrix $\overset{\leftrightarrow}{G}^q$ , where $1 \le t \le n\,, 0 \le q \le g-1$.We show how to compute Hamming weight of $\overset{\leftrightarrow}{G}^q$ as follows:

By $\overset{\leftrightarrow}{G}^q = \overbrace{\overset{\leftrightarrow}{B}_0 \circ \cdots \circ \overset{\leftrightarrow}{B}_0}^{g-q-1} \circ \overbrace{\overset{\leftrightarrow}{B}_1 \circ \cdots \circ \overset{\leftrightarrow}{B}_1}^{q}$ , then

$$H(\overset{\leftrightarrow}{G}^q)\big|_t = H(\overbrace{\overset{\leftrightarrow}{B}_0 \circ \cdots \circ \overset{\leftrightarrow}{B}_0}^{g-q-1} \circ \overbrace{\overset{\leftrightarrow}{B}_1 \circ \cdots \circ \overset{\leftrightarrow}{B}_1}^{q})\big|_t = \overbrace{H(\overset{\leftrightarrow}{B}_0)\big|_t \circ \cdots \circ H(\overset{\leftrightarrow}{B}_0)\big|_t}^{g-q-1} \circ \overbrace{H(\overset{\leftrightarrow}{B}_1)\big|_t \circ \cdots \circ H(\overset{\leftrightarrow}{B}_1)\big|_t}^{q}$$

$$= \overbrace{H(\overset{\leftrightarrow}{B}_0)\big|_t + \cdots + H(\overset{\leftrightarrow}{B}_0)\big|_t}^{g-q-1} + \overbrace{H(\overset{\leftrightarrow}{B}_1)\big|_t + \cdots + H(\overset{\leftrightarrow}{B}_1)\big|_t}^{q} .$$

When $g = 2$, $H(V^q)$ is a scalar and it reduces to the Hamming weight defined in Ref. [1].

***Theorem 3.1:*** Construction 3.1 above is a (*k*, *n*)-GVCS. The pixel expansion is $m_g = (g-1)m$ and the contrast $\alpha^{(q+1,q)}$ between $q+1$th and $q$-th are

$$m_g = (g-1)\, m\,,$$

$$\alpha^{(q+1,q)} = \frac{\alpha}{g-1} = \frac{h-l}{(g-1)m}\,, \quad q = 0, \cdots, g-2\,.$$

Notice that the two parameteres $m$ and $\alpha$ in the theorem above are pixel expansion and contrast of a binary $(k, n)$-nRVCS scheme.

***Proof:***

**To show the pixel expansion**, the pixel expansion $m_g = (g-1)\cdot m$ is obvious from the Construction 3.1 above.

**To show security**, we will prove that fact that $H(OR(\overleftrightarrow{G}^{q+1}|t)) = H(OR(\overleftrightarrow{G}^{q}|t))$ for $t \le k-1$. From the construction of the shares given in the section 3.3, we can see that the $(g-1)$ random matrices $\overleftrightarrow{G}^q$, which are $g-q-1$ $\overleftrightarrow{B}_0$ and $q$ $\overleftrightarrow{B}_1$, are all distinct and all independent of each other. Since $\overleftrightarrow{B}_0$ and $\overleftrightarrow{B}_1$ is the basis matrices of a $(n,k)$-VCS, according to the condition D-2 of definition 2.1 ( see section 2), we have $H(OR(B_1|t)) = H(OR(B_0|t))$ for $t \le k-1$.

As we know, in Boolean matrix $\overleftrightarrow{G}^q$ ( where $0 \le q \le g-1$), Columns of each Component $\overleftrightarrow{B}_0$ (resp. $\overleftrightarrow{B}_1$) or $\overleftrightarrow{B}_1$ (resp. $\overleftrightarrow{B}_0$) are randomly permutated within that Component independently, no any information can be obtained if less than $k$ share is stacked together. So, it is easy to verify that $H(OR(\overleftrightarrow{G}^{q+1}|t)) = H(OR(\overleftrightarrow{G}^{q}|t))$ for $t \le k-1$.With fewer than $k$ shares, no information about the secret image is revealed in Boolean matrix $\overleftrightarrow{G}^q$, thus the security of the system is ensured.

**To show contrast**, let $Y = \{r_1, \cdots, r_t\} \subset \{1, \cdots, n\}$ be a subset of any $t \ge k$ rows in an $n \times m_g$ matrix $\overleftrightarrow{G}^q$, and let $S^q|t$ be the $t \times m_g$ matrix that results from considering only those row in $Y$. The Hamming distance between $S^q|t$ and $S^{q+1}|t$ for $q \in [0, g-2]$ is

$$\alpha^{(q+1,q)} \cdot (g-1)m = \overbrace{H(B_0|t)+\cdots+H(B_0|t)}^{g-(q+1)} + \overbrace{H(B_1|t)+\cdots+H(B_1|t)}^{q} -$$

$$\overbrace{H(B_0|t)+\cdots+H(B_0|t)}^{g-q} + \overbrace{H(B_1|t)+\cdots+H(B_1|t)}^{q-1}$$

$$\alpha^{(q+1,q)} \cdot (g-1)m = (g-q-1)\ H(B_0|t) + (q)\ H(B_1|t) - (g-q)\ H(B_0|t) - (q-1)\ H(B_1|t)$$

$$\alpha^{(q+1,q)} \cdot (g-1)m \ = H(B_1|t) - H(B_0|t) = h-l$$

$$\alpha^{(q+1,q)} = \frac{h-l}{(g-1)m} = \frac{\alpha}{g-1}. \square$$

Observe that ***Theorem 3.1*** arrived at a conclusion that is the same as the one in [13, 14]. By Lemma 3.1, our scheme has minimum pixel expansion.

The construction 3.1 seems a minor improvement to the existing construction [13, 14], but it has powerful function and can be easily used to construction grayscale (*k*, *n*)-VCS with reversing.

## 3.4 Optimal contrast grayscale (*k*, *n*)-RGVCS for nPBVCS

Based on some of the ideas from Cimoto's binary RVCS, in our scheme, for each pixel of the original image and for each participant $i\ (1 \le i \le n)$, the dealer generates the corresponding block pixels, which involves $g-1$ pixels, in each transparency $Bt_{i,1}, \ldots, Bt_{i,m}$. In the reconstruction phase, any $k\ (k \le n)$ participants can reconstruct the greyscale image with optimal contrast by performing a sequence of stacking and reversing operations on their transparencies. The proposed scheme is described in Table 3.5. We use symbol "$\oplus$" to represent OR operation.

Table 3.5 Distribution phase and reconstruction phase of the proposed schme

| **Distribution phase** | | **Reconstruction phase** |
|---|---|---|
| **Step 1:** Chooses $g$ grey levels $S^q \in \overleftrightarrow{G}^q$, where $q = 0, \cdots, g-1$.<br>**Step 2**: For each participant $i$, consider the $(g-1)m$ string bits $s_{i,1}, \ldots, s_{i,m(g-1)}$ composing the ith row of $S^q$ and for the *r*-th run , put a $(g-1)$ pixels $Bs_{i,r} = (s_{i,r}, s_{i,m+r}, \cdots, s_{i,(g-2).m+r})$ on the transparency $Bt_{i,r}$ ,where $i = 1, \ldots, n$ , $r = 1, \ldots, m$ . | | Any $k$ participants $\{j_1, j_2, \cdots, j_k\} \subset \{1, \cdots, n\}$ reconstruct the secret image by computing:<br>**Step 1**: For *r*-th run, $r = 1, \ldots, m$.<br>$T_r = OR(Bt_{j_1,r}, Bt_{j_2,r}, \cdots, Bt_{j_k,r})$,<br>$= Bt_{j_1,r} \oplus Bt_{j_2,r} \oplus \cdots \oplus Bt_{j_k,r}$.<br>**Step 2:** $\overline{T}_r$ , $r = 1, \ldots, m$.<br>**Step3:** $U = \overline{T}_1 \oplus \cdots \oplus \overline{T}_m$<br>**Step4:** $\tilde{P} = \overline{U} = \overline{\overline{T}_1 \oplus \cdots \oplus \overline{T}_m}$ , which is the reconstructed secret image. |

***Theorem 3.2:*** Construction (see Table 3.5) above is a reversing based $(k,n)$-GVCS. The pixel expansion is $m_g = g-1$ , the contrast between $q+1$ *-th* level and $q$ *-th* level $\alpha^{q+1,q} = \frac{1}{g-1}$, $q = 0, \cdots, g-2$.

***Proof:***

From construction process above, we can see that $Bs_{i,r}$ is a $g-1$ dimension Boolean vector, so its pixel expansion is $g-1$.

**To show security,**

For $r$-run $(k,n)$-RGVCS, the first concern is that one should not get any secret information from his shares $Bt_{i,r}$ ,where $i = 1, \ldots, n$ , $r = 1, \ldots, m$ . Then we must prove the fact that

$H(Bt_{i,r}) = H(Bt_{i,r-1})$ for $1 \le r \le n-1$.

***Case 1:*** one should not get any secret information from his shares

Our scheme uses the concept of probabilistic scheme and delivers the elements in one row to shadows of different runs. In the same position of $(g-1)\cdot m$ different shadows, because the $\overset{\leftrightarrow}{B_0}$ and $\overset{\leftrightarrow}{B_1}$ are a binary $(k,n)$-nRVCS, thus it satisfies $H(B_0|t)=H(B_1|t)$ for any $t$ row, where $1\le t\le n$. From $\overset{\leftrightarrow}{G^q}=\overbrace{\overset{\leftrightarrow}{B_0}\circ\cdots\circ\overset{\leftrightarrow}{B_0}}^{g-q-1}\circ\overbrace{\overset{\leftrightarrow}{B_1}\circ\cdots\circ\overset{\leftrightarrow}{B_1}}^{q}$, where $0\le q\le g-1$, we obtain $H(Bt_{i,r})=H(Bt_{i,r-1})$ for $1\le r\le n-1$. More, there is no any mutual information among their own shadows. Therefore, the schemes satisfy the first security concern.

***Case 2:*** $H(Bt_{i,r})=H(Bt_{i,r-1})$ for $1\le r\le n-1$.

The $g$ random matrices $\overset{\leftrightarrow}{G^q}$, which are $g-q-1$ $\overset{\leftrightarrow}{B_0}$ and $q$ $\overset{\leftrightarrow}{B_1}$, are all distinct and all independent of each other. We know that $H(OR(B_1|t))=H(OR(B_0|t))$ for $t\le k-1$, from $Bs_{i,r}=(s_{i,r},s_{i,m+r},\dots,s_{i,(g-2)\cdot m+r})$, each element of $Bs_{i,r}$ is from different column of basis matrices $\overset{\leftrightarrow}{B_0}$ and $\overset{\leftrightarrow}{B_1}$, so $H(OR(Bt_{i,r})|t)$ for $\overset{\leftrightarrow}{G^q}$ is the same as $H(OR(Bt_{i,r})|t)$ for $\overset{\leftrightarrow}{G^{q+1}}$, where $q=0,\cdots,g-1$. So no any information can be obtained if less than $k$ shares are stacked together, thus the security of the system is ensured.

**To show contrast**,

For *r*-th run, $T_r=OR(Bt_{j_1,r},Bt_{j_2,r},\cdots,Bt_{j_k,r})$, $\{j_1,j_2,\cdots,j_k\}\subset\{1,\cdots,n\}$ , $r=1,\dots,m$.

$T_r=Bt_{j_1,r}\oplus Bt_{j_2,r}\oplus\cdots\oplus Bt_{j_k,r}$. Let $T_r=(a_{r,1},a_{r,2},\dots,a_{r,(g-1)})$, $r=1,\dots,m$.

$$\tilde{P}=\overline{U}=\overline{\overline{T_1}\oplus\cdots\oplus\overline{T_m}}=(a_{1,1},\cdots,a_{1,(g-1)})\times\cdots\times(a_{m,1},\cdots,a_{m,(g-1)})\,. \tag{3.1}$$

By $\overset{\leftrightarrow}{G^q}=\overbrace{\overset{\leftrightarrow}{B_0}\circ\cdots\circ\overset{\leftrightarrow}{B_0}}^{g-q-1}\circ\overbrace{\overset{\leftrightarrow}{B_1}\circ\cdots\circ\overset{\leftrightarrow}{B_1}}^{q}$, where $0\le q\le g-1$.

From (3.1), we calculate expectation of the reconstructed pixel ($\tilde{P}$) in $\overset{\leftrightarrow}{G^q}$.

$$E\left[H(\tilde{P}^q)\right]=E(H((a_{1,1},\cdots,a_{1,(g-1)})\times\cdots\times(a_{m,1},\cdots,a_{m,(g-1)})))$$

$$=\sum_j^m \Pr((a_{1,1},\cdots,a_{1,(g-1)})=\cdots=(a_{m,1},\cdots,a_{m,(g-1)})=1)$$

$$=\sum_j^m \Pr((a_{1,1},\cdots,a_{1,(g-1)})=1)\times\cdots\times\Pr((a_{m,1},\cdots,a_{m,(g-1)})=1) \tag{3.2}$$

According to the property of *PBVCS* [3-4], when $(a_{1,1},\cdots,a_{m,1})$ comes from any *k* of *n* rows of $B_0$, at least one element of $(a_{1,1},\cdots,a_{m,1})$ is "0", thus $a_{1,1}\times\cdots\times a_{m,1}$=0. When $(a_{1,1},\cdots,a_{m,1})$ comes from any *k* of *n* rows of $B_1$, each element of $(a_{1,1},\cdots,a_{m,1})$ is "1", thus $a_{1,1}\times\cdots\times a_{m,1}$=1.

From (3.2), we obtain

$$E\left[H(\tilde{P}^q)\right]=\sum_{j}^{m}\frac{q-1}{m(g-1)}=\frac{q}{g-1},\text{where } q=0,\ldots,g-1.$$

$$\alpha^{q+1,q}=E\left[H(\tilde{P}^{q+1})\right]-E\left[H(\tilde{P}^{q})\right]=\frac{q+1}{g-1}-\frac{q}{g-1}=\frac{1}{g-1}.\square$$

Obviously, when $g$ = 2, this scheme is equivalent to binary $(k,n)$-Cimato's scheme [6].

**The complexity of the reconstruction phase:**

It needs $m\cdot(g-1)\cdot(k-1)$ OR operations to obtain $T_r$ in $r=m$ runs. Then $m-1$ OR operations and $m+1$ NOT operations are required to get the reconstructed image. So the total operations are $(m(g-1)(k-1)+m-1)$ *ORs and* $(m+1)$ *NOTs*. Each participant holds $m$ shares. The size of shares becomes $(g-1)\cdot m$ times larger than that of the original secret image. The size of reconstructed image is $(g-1)$ time larger than that of the original secret image. While $m$ is fixed, and $g$ is constant, then the complexity of reconstruction is $O(k-1)$.

# 4. Optimal contrast grayscale reversing-based VCS for nPBVCS

Most reversing-based VCSs are based on perfect black VCS (PBVCS). Yang et al. [8-9] first gave a reversing-based scheme for non-perfect black VCS (nPBVCS). In this section, based on the proposed nR-WBCP-GVCS in section 3.4 above, we propose a corresponding reversing-based grayscale NPBVCS with optimal contrast.

Notice that Yang et al.'s binary NPBRVCS [8-9] cannot be directly used to perform grayscale reversing-based VCS, more detailed analysis of a three grey levels (2, 3)-GVCS is given in Appendix C.

As we know XOR operation in real number-space $R^n$ is commonly used to design some schemes. The definition of XOR product of two vectors or multi-vectors is not discussed in VCS, whereas some researches use the XOR product of vectors in these contents [7-9] without giving the definition. Next we give the formal definition of XOR product of two vectors or multi-vectors and list some properties in nVCS.

In the Boolean-space $B_2$, the standard XOR product of two vectors in nVCS is defined by $x \otimes y = (\xi_1 \otimes \eta_1, \cdots, \xi_n \otimes \eta_n)$, where $x = (\xi_1, \cdots, \xi_n)$ and $y = (\eta_1, \cdots, \eta_n)$. An XOR operation product in a Boolean vector space $B_2$ is a *bilinear function*, the following proposition gives its properties.

***Proposition 4.1:***

(i) $x \otimes y = y \otimes x$.

(ii) $x \otimes x \geq 0$, and $x \otimes x = 0$ only for vector $x = 0$.

(iii) $x \otimes (y + z) = x \otimes y + x \otimes z$.

The proof is simple, and is omitted. The XOR operation product of two vectors can be expanded to that of multi-vectors. In fact, three properties of **Proposition 4.1** above satisfy XOR operation product of multi-vectors.

We now make minor change to $\overset{\leftrightarrow}{G^q} = \overbrace{\overset{\leftrightarrow}{B_0} \circ \cdots \circ \overset{\leftrightarrow}{B_0}}^{g-q-1} \circ \overbrace{\overset{\leftrightarrow}{B_1} \circ \cdots \circ \overset{\leftrightarrow}{B_1}}^{q}$ (see **Construction 3.1**), and then we obtain grey levels basis matrices $\Gamma(\overset{\leftrightarrow}{G^q})$ for $(k, n)$-RGVCS with cyclic-shift operations as follows.

$$\overleftrightarrow{\Gamma}(G^q) = \overbrace{\overleftrightarrow{\Gamma}(B_0)\circ\cdots\circ\overleftrightarrow{\Gamma}(B_0)}^{g-q-1}\circ\overbrace{\overleftrightarrow{\Gamma}(B_1)\circ\cdots\circ\overleftrightarrow{\Gamma}(B_1)}^{q},\quad 0\le q\le g-1.$$

The cyclic-shift operation of $\overleftrightarrow{\Gamma}(G^q)$ only performs local column cyclic-shift move. The local cyclic-shift operation is that columns of each Component are cyclic-shift operations moved within that Component (such as $\overleftrightarrow{\Gamma}(B_0)$ or $\overleftrightarrow{\Gamma}(B_1)$), no columns of different Components are moved, and all Components go through exactly the same internal column cyclic-shift operation simultaneously.

The cyclic-shift operation of $\overleftrightarrow{\Gamma}(B_0)$ (resp. $\overleftrightarrow{\Gamma}(B_1)$) is $\Gamma(B_0)(reps.\Gamma(B_1)) = [\gamma(s_{i,j_k})]$, where $\gamma(\cdot)$ is a 1-bit cyclical right shift function. i.e. $\gamma(s_{ij_1}...s_{ij_m}) = (s_{ij_m}s_{ij_1},...s_{ij_{m-1}})$.

Based on the above discussion, we now propose a reversing-based greyscale $(k,n)$-NPBVCS. The distribution phase and reconstruction phase are given in Table 4.1.

Table 4.1 The proposed reversing-based greyscale $(k,n)$ - nPBVCS

| **Distribution phase** | **Reconstruction phase** |
|---|---|
| Grey level palette $\Gamma_{Grey} = \{i_0,\dots,i_{g-1}\}$, $g\ge 1$, a family of $g$ collections of $n\times m_g$ Boolean matrices $\{BS^0, BS^1,\cdots,BS^{g-1}\}$ where $BS^q \in \overleftrightarrow{\Gamma}(G^q)$ is the collection for grey level $i_q$ for $0\le q\le g-1$. The dealer,<br>**Step1:** Chooses $g$ grey levels $BS^q \in \overleftrightarrow{\Gamma}(G^q)$, where $q=0,\cdots,g-1$. And performs an $(k,n)$-scheme to generate $n$ shadows $A_1^1,\dots,A_n^1$ to $n$ participants for the first run.<br>**Step2:** Generates the shadows $A_j^r = \Gamma(A_j^{r-1})$ to $n$ participants for $r$-th round, $j=1,\cdots,n$, $r=2,\cdots,m$. Note the shadow should be labeled as to which run it is, for easy management by the participant. | Any $k$ participants reconstruct the secret image by next steps.<br>**Step 1:** To recover the secret within $m$ runs, at least $k$ participants, participants $j_1,\dots,j_k$, offer their $(k\times m)$ shadows $A_{j_1}^r,\dots,A_{j_k}^r$, $r\in[1,m]$, for reconstruction, $j_1,\dots,j_k \in \{1,\dots,n\}$, $j_1\ne\cdots\ne j_k$.<br>**Step 2:** Stack the shadows $A_{j_1}^r,\dots,A_{j_k}^r$ to reconstruct the image $GT_r$ in the $r$-th run. $GT_r = A_{j_1}^r \oplus A_{j_2}^r \oplus\dots\oplus A_{j_k}^r$, $r=1,\dots,m$,.<br>**Step 3:** Finish $m$ runs by using XOR operation to reconstruct $U^q = GT_1\otimes\cdots\otimes GT_m$.<br>**Step 4:** If '$m-h$' is even (i.e. '$m-h$' is odd) then the reconstructed image $\tilde{P}$ is $\tilde{P}=U^q$; otherwise the reconstructed image is $\tilde{P}=\overline{U^q}$. |

***Theorem 4.1:*** If $(m-h)$ is even positive integer and $(m-l)$ is odd positive integer, construction 4.1 above is a reversing based proposed $(k,n)-GVCS$ above with pixel expansion $m_g = m\cdot(g-1)$ and the contrast difference between $q+1$-th level and $q$-th level $\alpha^{q+1,q} = \dfrac{1}{g-1}, q=0,\cdots,g-2$.

***Proof:***

**To show security,**

In first run, the dealer employs the $(k,n)$-GVCS proposed in section 3.4 to create the $n$ shares (or shadows) $A_1^1,\ldots,A_n^1$ to $n$ participants. We obtain $H(OR(\overset{\leftrightarrow}{G}^{q+1}|t)) = H(OR(\overset{\leftrightarrow}{G}^{q}|t))$ for $t \le k-1$, where $0 \le q \le g-1$. So with fewer than $k$ shares, no information about the secret image is revealed in $A_j^1$ , thus the security of the system is ensured in the first run.

Then, the dealer performs shift operation on these $n$ shadows to generate $m-1$ runs. Namely, generate the shares $A_j^r = \Gamma(A_j^{r-1})$ to $n$ participants for rth round, $j=1,\cdots,n$ , $r = 2,\cdots,m$ . For $t \le k-1$ , each participant $j$ holds $m-1$ shares, which are obtained by performing the shift operation on corresponding $A_j^1$ in first run.

It is clearly that $H(OR(\overset{\leftrightarrow}{G}^{q+1}|t)) = H(OR(\overset{\leftrightarrow}{G}^{q}|t))$ for $r$th round. According to the security of the proposed GVCS, the scheme satisfies the security property.

With fewer than $k$ shares, no information about the secret image is revealed in $Bs_j^r$ , thus the security of the scheme is ensured.

**To show contrast,**

From the reconstruction process above we obtain

$$U^q|t = (GT_1 \otimes \cdots \otimes GT_m)|t = (GT_1)|t \otimes \cdots \otimes (GT_m)|t \text{, where } t \ge k. \quad (4.1)$$

Form basis matrix above, $\Gamma(\overset{\leftrightarrow}{G}^q) = \overbrace{\Gamma(\overset{\leftrightarrow}{B}_0)\circ\cdots\circ\Gamma(\overset{\leftrightarrow}{B}_0)}^{g-q-1}\circ\overbrace{\Gamma(\overset{\leftrightarrow}{B}_1)\circ\cdots\circ\Gamma(\overset{\leftrightarrow}{B}_1)}^{q}$, $0 \le q \le g-1$.

Then $$GT_1|t = \overbrace{OR(\overset{\leftrightarrow}{B}_0|t)\circ\cdots\circ OR(\overset{\leftrightarrow}{B}_0|t)}^{g-q-1}\circ\overbrace{OR(\overset{\leftrightarrow}{B}_1|t)\circ\cdots\circ OR(\overset{\leftrightarrow}{B}_1|t)}^{q} \quad (4.2)$$

Since $B_0$ and $B_1$ are basis matrices of a binary $(k,n)$ -*nVCS,* thus $OR(B_0|t)$ includes '$h$'W '$(m-h)$' B, $OR(B_1|t)$ has '$l$' W '$(m-l)$' B (see condition (D-1) in section 2.1 ).

For convenience, we use vector $(t_1^0,\cdots,t_m^0)$ to represent vector of $OR(B_0|t)$, and vector $(t_1^1,\cdots,t_m^1)$ to represent the vector of $OR(B_1|t)$.

By (4.2), we obtain

$$GT_1|t = \overbrace{(t_1^0,\cdots,t_m^0)\circ\cdots\circ(t_1^0,\cdots,t_m^0)}^{g-q-1}\circ\overbrace{(t_1^1,\cdots,t_m^1)\circ\cdots\circ(t_1^1,\cdots,t_m^1)}^{q}. \quad (4.3)$$

$$GT_2|t = \Gamma(GT_1|t) = \overbrace{\Gamma(t_1^0,\cdots,t_m^0)\circ\cdots\circ\Gamma(t_1^0,\cdots,t_m^0)}^{g-q-1}\circ\overbrace{\Gamma(t_1^1,\cdots,t_m^1)\circ\cdots\circ\Gamma(t_1^1,\cdots,t_m^1)}^{q}$$

$$GT_r|t = \Gamma(GT_{r-1}|t),\ r = 3,...,m. \tag{4.4}$$

Applying result of **Proposition 4.1**, and as we know the "$\circ$" operation, which is Concatenation operation, and "$\otimes$"operation satisfy commutative law in a real number space. By (4.1), we get that

$$U^q|t = \overbrace{(t_1^0\otimes\cdots t_m^0,\cdots,t_1^0\otimes\cdots t_m^0)\circ\cdots\circ(t_1^0\otimes\cdots t_m^0,\cdots,t_1^0\otimes\cdots t_m^0)}^{g-q-1}\circ$$

$$\circ\overbrace{(t_1^1\otimes\cdots t_m^1,\cdots,t_1^1\otimes\cdots t_m^1)\circ\cdots\circ(t_1^1\otimes\cdots t_m^1,\cdots,t_1^1\otimes\cdots t_m^1)}^{q} \tag{4.5}$$

We know the $t_1^0\otimes\cdots t_m^0$ (resp. $t_1^1\otimes\cdots t_m^1$ ) is from $OR(B_0|t)$ $(resp. OR(B_1|t)$ , which includes '$h$' W '$(m-h)$' B (resp. '$l$' W'$(m-l)$'B) (see definition 2.1).

If $(m-h)$ is even positive integer and $(m-l)$ is odd positive integer, thus $t_1^0\otimes\cdots t_m^0$=0 and $t_1^1\otimes\cdots t_m^1$=1.

From (4.5), we obtain that

$$U^q|t = \overbrace{\overbrace{0,\cdots,0}^{m},\cdots,\overbrace{0,\cdots,0}^{m}}^{g-1-q},\overbrace{\overbrace{1,\cdots,1}^{m},\cdots,\overbrace{1,\cdots,1}^{m}}^{q}$$

$$H(U^q|t) = H(\overbrace{\overbrace{0,\cdots,0}^{m},\cdots,\overbrace{0,\cdots,0}^{m}}^{g-1-q},\overbrace{\overbrace{1,\cdots,1}^{m},\cdots,\overbrace{1,\cdots,1}^{m}}^{q}) = q\cdot m \tag{4.6}$$

If $(m-h)$ is odd positive integer and $(m-l)$ is even positive integer, thus $t_1^0\otimes\cdots t_m^0$=1 and $t_1^1\otimes\cdots t_m^1$=0.

$$U^q|t = \overbrace{\overbrace{1,\cdots,1}^{m},\cdots,\overbrace{1,\cdots,1}^{m}}^{g-1-q},\overbrace{\overbrace{0,\cdots,0}^{m},\cdots,\overbrace{0,\cdots,0}^{m}}^{q}$$

$$H(U^q|t) = H(\overbrace{\overbrace{1,\cdots,1}^{m},\cdots,\overbrace{1,\cdots,1}^{m}}^{g-1-q},\overbrace{\overbrace{0,\cdots,0}^{m},\cdots,\overbrace{0,\cdots,0}^{m}}^{q}) = (g-1-q)\cdot m \tag{4.7}$$

From (4.6), the contrast $\alpha^{q+1,q}$ is

$$\alpha^{q+1,q} = \frac{H(U^{q+1}|t)-H(U^q|t)}{(g-1)m} = \frac{m\cdot(q+1)-m\cdot q}{(g-1)m} = \frac{1}{g-1},\ q = 0,\cdots,g-2$$

From (4.7),

Contrast $\alpha^{q+1,q} = \frac{H(\bar{U}^{q+1}) - H(\bar{U}^{q})}{(g-1)m} = \frac{-m\cdot(g-1-a-1)+m\cdot(g-1-q)}{(g-1)m} = \frac{1}{g-1}$, $q = 0,\cdots,g-2$ .□

In Theorem 4.1 above, when $g=2$, this scheme is equivalent to the $(k,n)$-NPBRVCS scheme in [8-9] .

**The decoding complexity of the theorem 4.1:** It needs $m(k-1)$ OR operations to obtain $GT_r$ in $r(=1,\dots,m)$ runs. Then $m-1$ XOR and one NOT are required to finish $m$ runs. So the total operations are $(m(k-1)+3(m-1))=(mk+2m-3)\ ORs\ and\ (4(m-1)+1)\ NOTs$. The number of shares held by each participant is $m^2$. The size of reconstructed image is $(g-1)$ times larger than that of the original secret image. Since $m$ is fixed, and $g$ is constant, then the complexity of reconstruction is $O(k)$.

## 5. Optimal contrast greyscale RVCS with minimum number of shares

Inspired by Hu and Tzeng's binary RVCS [7], we first use matrix concatenation to construct basis matrices and an **Open auxiliary matrix** for a $(k,n)$-GVCS with $g$ grey levels, then we propose a reversing-based optimal-contrast grayscale VCS (RGVCS).

### 5.1 The proposed grayscale $(k,n)$ -nRVCS using $(k,k)$ -nRGVCS

We now give a basis and Auxiliary matrix $(k,n)$-nRVCS using $(k,k)$-nRVCS by the following steps.

---

**Construction 5.1: Construct basis and Auxiliary matrix for a $(k,n)$-GVCS with g grey levels**

***Input:*** $G^q$ Basis matrix of a $g$ grey levels $(k,k)$-GVCS with $m_g$, $0 \le q \le g-1$.

***Output:*** $L^q$ Basis matrix for a $(k,n)$-GVCS with $g \ge 2$ grey levels and its Auxiliary matrix.

***Matrix construction procedure:***

For basis matrix of a $(k,n)$ scheme, we create a construction matrix with $n$ rows from the $k$ rows of the construction matrix of the $(k,k)$- nRVCS scheme as described in [1]. We do it in five steps.

***Step 1***: Generate $v=\binom{n}{k}$ distinct construction matrices for $\binom{n}{k}$ different ( $(k,k)$ -GVS schemes to the same secret image, namely, $Ge_p^q$, $p=1,\cdots,v$. Here, we denote $v$ be the number of $k$-combinations of an $n$-element set.

***Step 2:*** Consider a function $f: Z^+ \to Z^+$, $q \in \{1,\cdots,k\}$, $f(q) \in \{1,\cdots,n\}$ ,for example, when $n=3$ and $k=2$ ,one possible such functions are $f(1)=1, f(2)=2$ ,or $f(2)=1, f(3)=2$, or $f(1)=1, f(3)=2$. There are v different ways to define such a function. Let $w \in \{1,\cdots,v\}$ and $l_w$ be one of such functions.

***Step 3:*** Generate a random matrix $L^q$ of $n$ rows, $\overset{\leftrightarrow}{Ge}{}_p^{(k,n)} = \begin{bmatrix} V_1^{(w)} \\ \vdots \\ V_n^{(w)} \end{bmatrix}$. For $q \in \{1,\cdots,k\}$, set $V_{q'}^{(w)} = V_q^{(w)}$ and $q' = f_w(q)$. In other words, substitute $k$ rows of $Ge_p^{(k,n)}$ with the rows of $G^q$ according to function $f_w$. For example, with $n=3$ and $k=2$, $\overset{\leftrightarrow}{Ge}{}_p^{(k,n)}$ could be $\begin{bmatrix} V_1^{(1)} \\ V_2^{(1)} \\ r \end{bmatrix}$, or $\begin{bmatrix} r \\ V_1^{(2)} \\ V_2^{(2)} \end{bmatrix}$, or $\begin{bmatrix} V_1^{(3)} \\ r \\ V_2^{(3)} \end{bmatrix}$, where $r$ is row vector with full 1's.

***Step 4:*** Concatenate all v different matrices $Ge_p^{(k,n)}$ together and obtain $L^q = \overset{\leftrightarrow}{Ge}{}_1^{(k,n)} \circ \overset{\leftrightarrow}{Ge}{}_2^{(k,n)} \circ \cdots \circ \overset{\leftrightarrow}{Ge}{}_v^{(k,n)}$ as the resulting $n \times (m \cdot v)$ .Construction matrix for our $(k,n)$ scheme. Notice that each $Ge_p^{(k,n)}$ is different from $G^q$.

---

*Step 5:* The $g$ grey levels **Open auxiliary matrix** $GA$ is the same matrix as $L^q$ except that we replace all the elements of the corresponding that of $(k,k)-nRGVCS$ with all 0's, $\overleftrightarrow{GA} = \overleftrightarrow{FF}_1 \circ \cdots \circ \overleftrightarrow{FF}_v$ .

We give example 5.1 to illustrate the construction above.

***Example 5.1:*** The greyscale (2, 3)- nRGVCS scheme with $g = 3$.

The basis matrices of (2, 2) -nRGVCS are $Ge^0 = \begin{bmatrix}1010\\1010\end{bmatrix}$, $Ge^1 = \begin{bmatrix}1010\\1001\end{bmatrix}$, $Ge^2 = \begin{bmatrix}1010\\0101\end{bmatrix}$.

Using the $\binom{3}{2}$ possible functions $f$, we create 3 matrices $Ge_p^{(k,n)}$ as follows:

$$\overleftrightarrow{Ge}_1^{0(1,2)} = \begin{bmatrix}\overbrace{1010}^{\{1,2\}}\\1010\\1111\end{bmatrix}, \overleftrightarrow{Ge}_2^{0(1,3)} = \begin{bmatrix}\overbrace{1010}^{\{1,3\}}\\1111\\1010\end{bmatrix}, \overleftrightarrow{Ge}_3^{0(2,3)} = \begin{bmatrix}\overbrace{1111}^{\{2,3\}}\\1010\\1010\end{bmatrix},$$

$$\overleftrightarrow{Ge}_1^{1(1,2)} = \begin{bmatrix}\overbrace{1010}^{\{1,2\}}\\1001\\1111\end{bmatrix}, \overleftrightarrow{Ge}_2^{1(1,3)} = \begin{bmatrix}\overbrace{1010}^{\{1,3\}}\\1111\\1001\end{bmatrix}, \overleftrightarrow{Ge}_3^{1(2,3)} = \begin{bmatrix}\overbrace{1111}^{\{2,3\}}\\1010\\1001\end{bmatrix},$$

$$\overleftrightarrow{Ge}_1^{2(1,2)} = \begin{bmatrix}\overbrace{1010}^{\{1,2\}}\\0101\\1111\end{bmatrix}, \overleftrightarrow{Ge}_2^{2(1,3)} = \begin{bmatrix}\overbrace{1010}^{\{1,3\}}\\1111\\0101\end{bmatrix}, \overleftrightarrow{Ge}_3^{2(2,3)} = \begin{bmatrix}\overbrace{1111}^{\{2,3\}}\\1010\\0101\end{bmatrix}.$$

The first two rows of $\overleftrightarrow{Ge}_p^{q(2,3)}$ are from the first two $Ge^q$ matrices. The first row and the third row of $\overleftrightarrow{Ge}_p^{q(2,3)}$ are from the first row and the second row of $Ge^q$. The second row and the third row of $\overleftrightarrow{Ge}_p^{q(2,3)}$ are from the first row and the second row of $Ge^q$. Other rows of $\overleftrightarrow{Ge}_p^{q(2,3)}$ are full 1.

In matrix $L^q$, the concatenation of these $\binom{3}{2}$ matrices forms the basic matrix as below.

$$L^0 = \overleftrightarrow{Ge}_1^{0(1,2)} \circ \overleftrightarrow{Ge}_2^{0(1,3)} \circ \overleftrightarrow{Ge}_3^{0(2,3)}, L^1 = \overleftrightarrow{Ge}_1^{1(1,2)} \circ \overleftrightarrow{Ge}_2^{1(1,3)} \circ \overleftrightarrow{Ge}_3^{1(2,3)}, L^2 = \overleftrightarrow{Ge}_1^{2(1,2)} \circ \overleftrightarrow{Ge}_2^{2(1,3)} \circ \overleftrightarrow{Ge}_3^{2(2,3)}.$$

$$L^0 = \left[\begin{array}{cccc|cccc|cccc}1&0&1&0&1&0&1&0&1&1&1&1\\1&0&1&0&1&1&1&1&1&0&1&0\\1&1&1&1&1&0&1&0&1&0&1&0\end{array}\right], \quad L^1 = \left[\begin{array}{cccc|cccc|cccc}1&0&1&0&1&0&1&0&1&1&1&1\\1&0&0&1&1&1&1&1&1&0&1&0\\1&1&1&1&1&0&0&1&1&0&0&1\end{array}\right],$$

$$L^2 = \left[\begin{array}{cccc|cccc|cccc} 1 & 0 & 1 & 0 & 1 & 0 & 1 & 0 & 1 & 1 & 1 & 1 \\ 0 & 1 & 0 & 1 & 1 & 1 & 1 & 1 & 1 & 0 & 1 & 0 \\ 1 & 1 & 1 & 1 & 0 & 1 & 0 & 1 & 0 & 1 & 0 & 1 \end{array}\right], GA = \left[\begin{array}{cccc|cccc|cccc} 0 & 0 & 0 & 0 & 0 & 0 & 0 & 0 & 1 & 1 & 1 & 1 \\ 0 & 0 & 0 & 0 & 1 & 1 & 1 & 1 & 0 & 0 & 0 & 0 \\ 1 & 1 & 1 & 1 & 0 & 0 & 0 & 0 & 0 & 0 & 0 & 0 \end{array}\right].$$

## 5.2 **Optimal contrast grayscale (*k*, *n*)-RVCS with minimum number of shares held**

Using the scheme proposed in subsection 5.1, Table 5.2 gives the distribution phase and reconstruction phase for a (*k*, *n*)-RGVCS

Table 5.2 Distribution phase and reconstruction phase for a (*k*, *n*)-RGVCS

<table>
<tr><td colspan="2">Let $S^q$ be collect of basis matrix $L^q$, where $0 \le q \le g-1$. Let $B^0$ be the collection of Boolean matrix $GA$.</td></tr>
<tr><td>Distribution phase</td><td>Reconstruction phase</td></tr>
<tr><td>Encodes each share $t_i$ as $(k,n)$ sub-shares $t_i, p$ and each sub-block consists of one secret image. $p = 1,\cdots,v$, $v = \binom{n}{k}$. Each $q$ grey levels on sub-block $t_i, p$ is encoded using $\overset{\leftrightarrow}{Ge}_p^{(k,n)}$.To share a $q$ grey levels, the dealer,<br>Step 1: Chooses $g$ grey levels $S^q \in L^q$, where $q = 0,\cdots,g-1$ .For each participant $i$, put a $m_g \cdot v$ pixels $s_{i,1},\ldots,s_{i,m_g \cdot v}$ on the transparency $t_i$ for the 1st run, where $i = 1,\ldots,n$.<br>Step 2: Chooses Open auxiliary matrix $B^q \in GA$, for each participant $i$ put $(a_{i,1}, a_{i,2},\ldots,a_{i,m_g \cdot v})$ string bits on the transparency $A_j$ for 2nd run, $j = 1,\cdots,n$.</td><td>Any $k$ participants in $(k,n)$ scheme reconstruct the secret image by computing:<br>Step 1: XORing any $k$ of $n$ shares $t_i, A_j$, $T = t_{j_1} \otimes \cdots \otimes t_{j_k}$, $A = A_{j_1} \otimes \cdots \otimes A_{j_k}$, where $\{j_1, j_2, \cdots, j_k\} \subset \{1,\cdots,n\}$.<br>Step 2: Compute $U = (T \oplus A) \otimes A$, $\tilde{P} = U$ is the reconstructed secret image.</td></tr>
</table>

***Theorem 5.2:*** The algorithm proposed above is a $(k,n)-RGVCS$, pixel expansion $\bar{m}_g = m_g \cdot \binom{n}{k}$, the contrast difference between $q+1$-th level and $q$-th level $\bar{\alpha}^{q+1,q} = \frac{1}{g-1}$, $q = 0,\cdots,g-2$.

***Proof:***

***To show security***,

We need to prove that any $t$ rows in $L^q$ cannot obtain any information about the secret image, each $i$ row in $L^q$ cannot leak any information of the secret image, and any $t$ participants cannot also reconstruct the secret image from ***Open auxiliary matrix,*** here $t \le k-1$, $1 \le i \le n$. This can be proved by three parts as follows.

***(i) Part 1: we cannot get any information of secret image form any $t$ rows of basis matrix $L^q$.***

From the construction method above (see in 5.1), in $g$ grey levels matrix $L^q = \overset{\leftrightarrow}{Ge}_1^{(k,n)} \circ \overset{\leftrightarrow}{Ge}_2^{(k,n)} \circ \cdots \circ \overset{\leftrightarrow}{Ge}_v^{(k,n)}$ ,the shares $\overset{\leftrightarrow}{Ge}_1^{(k,n)}, \overset{\leftrightarrow}{Ge}_2^{(k,n)}, \cdots, \overset{\leftrightarrow}{Ge}_v^{(k,n)}$ are all random and all independent of each other. Each $\overset{\leftrightarrow}{Ge}_p^{(k,n)}$ $(1 \le p \le v = \binom{n}{k})$ comes different $(k,k)$-GVCS with the secret image.

For $t \le k-1$, $H(\overset{\leftrightarrow}{Ge}_1^{(k,n)} | t) = H(\overset{\leftrightarrow}{Ge}_2^{(k,n)} | t) = \cdots = H(\overset{\leftrightarrow}{Ge}_v^{(k,n)} | t)$,

then $H(\overset{\leftrightarrow}{L^0} | t) = H(\overset{\leftrightarrow}{L^1} | t) = \cdots = H(\overset{\leftrightarrow}{L^{g-1}} | t)$.

We see that any $k-1$ rows cannot recover any information about the secret image.

***(ii) Part 2: each*** $i$ ***row in*** $L^q$ ***cannot leak any information of the secret image.***

The matrix $\overset{\leftrightarrow}{Ge}_p^{(k,n)}$ is a special $(k,n)$-GVCS, which can construct the secret image using special $k$ rows of $n$ rows. In basis Boolean matrix $L^q$, each $i$ row of $L^q$ maybe includes $k$ block rows of $(k,k)$-GVCS, we know that the block $k$ rows of the $i$ row in matrix $L^q$ are from different $\overset{\leftrightarrow}{Ge}_p^{(k,n)}$ according to construction 5.1 above, we use independent randomly permutation on $\overset{\leftrightarrow}{Ge}_p^{(k,n)}$, so the $k$ block rows in the $i$ row of matrix $L^q$ is from $k$ rows of different $(k,k)$-GVCS. So the $k$ block rows cannot construct any information of the secret image. In matrix $\overset{\leftrightarrow}{Ge}_p^{(k,n)}$, there exist full 1 rows, which have not any contribution to recover secret image. So, we cannot get any information of the secret image from the special rows of the matrix $\overset{\leftrightarrow}{Ge}_p^{(k,n)}$.

***(iii) Part 3: the security of Open auxiliary matrix*** $GA$

From the construction method above (see in 5.1), as we know the ***Open auxiliary matrix*** $GA$ consists of rows and columns with full 1 and full 0, it is only marked position of secret share image, so the matrix $GA$ does not share any information of the secret image.

By three parts above, each row of the matrix $L^q$ is a random matrix. With fewer than $k$ shares of $L^q$, no information about the secret image is revealed. The ***Open auxiliary matrix*** $GA$ cannot share any information of the secret image, thus the security of the system is ensured.

***To show the pixel expansion***,

The pixel expansion $\bar{m}_g = m_g \cdot \binom{n}{k}$ is obvious from the shadow construction process above.

***To show contrast,***

Now we begin to compute the contrast of the recovered secret image when any $k$ participants perform XOR operations the $k$ shares and stacking all the shares in $(k,n)$ scheme.

Let $T_p$ (resp. $A_p$) represent the result of XORing any $k$ shares p-*th* block grey levels matrix $\overset{\leftrightarrow}{Ge}_p^{(k,n)}|k$ (resp. $\overset{\leftrightarrow}{FF}_p^{(k,n)}|k$), where $p = 1, \cdots, v$.

$$T_p = t_{p,j_1} \otimes \cdots \otimes t_{p,j_k}, A_p = A_{p,j_1} \otimes \cdots \otimes A_{p,j_k}, \{j_1, j_2, \cdots, j_k\} \subset \{1, \cdots, n\}, 1 \le p \le v.$$

Thus $T = T_1 \oplus T_2 \oplus \cdots \oplus T_v$, $A = A_1 \oplus A_2 \oplus \cdots \oplus A_v$.

$$\begin{aligned} H(U^q)|k &= H((T \oplus A) \otimes A)|k \\ &= H((T_1 \oplus A_1 \oplus T_2 \oplus A_2 \oplus \cdots T_v \oplus A_v) \otimes (A_1 \oplus A_2 \oplus \cdots A_v))|k \\ &= H((T_1 \oplus A_1) \otimes A_1)|k + H((T_2 \oplus A_2) \otimes A_2)|k + \cdots H((T_v \oplus A_v) \otimes A_v)|k \end{aligned} \quad (5.1)$$

As we know that the matrix $L^q$ $(= \overset{\leftrightarrow}{Ge}_1^{(k,n)} \circ \overset{\leftrightarrow}{Ge}_2^{(k,n)} \circ \cdots \circ \overset{\leftrightarrow}{Ge}_v^{(k,n)})$ includes $v (= \binom{n}{k})$ distinct sub-matrices $\overset{\leftrightarrow}{Ge}_1^{(k,n)}, \overset{\leftrightarrow}{Ge}_2^{(k,n)}, \cdots, \overset{\leftrightarrow}{Ge}_v^{(k,n)}$. In matrix $L^q$, there exist some special rows, which come from $v$ different matrix $G^q$ with secret image. From the construction method above (see in 5.1), when we fix $k$ row vectors $j_1, j_2, \cdots, j_k$ ($\{j_1, j_2, \cdots, j_k\} \subset \{1, \cdots, n\}$), in this case only $p$-th block matrix $\overset{\leftrightarrow}{Ge}_p^{(k,n)}|k$ and $\overset{\leftrightarrow}{FF}_p|k$ in $L^q$ and $\overset{\leftrightarrow}{GA} (= \overset{\leftrightarrow}{FF}_1 \circ \cdots \circ \overset{\leftrightarrow}{FF}_v)$ can construct the secret image, $p \in \{1, \cdots, v\}$.

Without loss of generality, suppose $p = 1$. In matrix $\overset{\leftrightarrow}{Ge}_1^{(k,n)}$, the previous $k$ row vectors are from $k$ rows of grey levels matrix $G^q$. Other rows are full 1 in $\overset{\leftrightarrow}{Ge}_1^{(k,n)}$. Because previous $k$ row vectors are full 0 in $\overset{\leftrightarrow}{FF}_1$ other rows are full 1, thus compute XOR operations corresponding $k$ row vectors, we obtain $A_1 = 0$.

When $p \ne 1$, there exists a row with full 1 elements in $\overset{\leftrightarrow}{FF}_p$, other rows are full 0, then we get $A_p = 1$, thus $H((T_2 \oplus A_2) \otimes A_2)|k = \cdots H((T_v \oplus A_v) \otimes A_v)|k = 0$.

By (5.1), we obtain that

$$H(U^q)|k = H((T \oplus A) \otimes A)|k = H((T_1 \oplus A_1) \otimes A_1)|k + H((T_2 \oplus A_2) \otimes A_2)|k + \cdots H((T_v \oplus A_v) \otimes A_v)|k$$

$$= H((T_1 \oplus A_1) \otimes A_1)|k = H((T_1 \oplus 0) \otimes 0)|k = H(T_1)|k \qquad (5.2)$$

Now that we recall the pervious $k$ row vectors in $\overset{\leftrightarrow}{Ge_1}^{(k,n)}$, which equal the $k$ rows is that of the matrix $G^q$. Thus $T_1$ is equal to XORing $k$ row vectors of $G^q$.

Because $B_0$ and $B_1$ are basis matrices of Naor and Shamir's $(k,k)$ scheme [1]. We know that $B_1$ ( reps. $B_0$ ) is the matrix whose columns are all the Boolean $k$-vectors having an odd (resp.even) number of 1's [1]. it is easy to verify that the white pixels are all white while $k$ participants perform XOR operations on the $k$ shares by computer $t_{p,j_1} \otimes \cdots \otimes t_{p,j_k}$, the black pixel are all black by computer $t_{p,j_1} \otimes \cdots \otimes t_{p,j_k}$.

Consider $G^q = \overbrace{B_0 \circ \cdots \circ B_0}^{g-q-1} \circ \overbrace{B_1 \circ \cdots \circ B_1}^{q}$, XORing operations on $k$ row vectors of $G^q$ are equal to performing XOR operations to $g-q-1$ basis matrices $B_0$ and performing XOR operations to $q$ basis matrices $B_1$. Let $m$ is pixel expansion of matrix $B_1$(or $B_0$).

By $G^q = \overbrace{B_0 \circ \cdots \circ B_0}^{g-q-1} \circ \overbrace{B_1 \circ \cdots \circ B_1}^{q}$, we obtain that

$$T_1 = t_{1,j_1} \otimes \cdots \otimes t_{1,j_k} = \overbrace{0 + \cdots + 0}^{g-q-1} + \overbrace{m + \cdots + m}^{q}, \quad q = 0, \cdots, g-1.$$

From (5.2), we get that

$$H(U^q)|k = H(T_1)|k = q \cdot m.$$

Then $\alpha^{q+1,q} = \dfrac{H(U^{q+1}) - H(U^q)}{m_g} = \dfrac{m(q+1-q)}{(g-1)m} = \dfrac{1}{g-1}$. □

When $g$ = 2, it can degenerate to a binary $(k,n)$-RVCS. From Theorem 5.2, we get the following corollary 5.1.

***Corollary 5.1:*** In a binary RVCS, contrast of the reconstructed black pixel and white pixel is

$$\alpha^{1,0} = \frac{1}{2-1} = 1 .$$

Obviously, when $g$ = 2, this scheme is equivalent to that of Hu and Tzeng's scheme [7] .

**The complexity of the reconstruction phase:** It needs $4k$ OR operations to obtain $T$ in 1 runs. Then $4k-1$ NOT operations are required to get the reconstructed image. So the total operations are $(4k)\ ORs\ and\ (4k-1)\ NOTs$ . Each participant holds 1 share. The size of shares becomes $\bar{m}_g$ times larger than that of the original secret image. The size of reconstructed image is $(g-1)$ times larger than that of the original secret image. While $m$ is fixed, and $g$ is constant, then the complexity of reconstruction is $O(k)$ .

# 6. Comparisons and discussions

In this subsection, we will compare the proposed schemes with grayscale nRVCS, and non-visual secret sharing schemes.

### 6.1 Comparison *with grayscale nRVCS*

Even if we do not have the copy machine with reversing operation, our schemes can reconstruct the secret image by stacking the shares directly. Our proposed schemes are fully compatible to the binary RVCSs, which are fully compatible to the traditional nRVCSs.

In a grayscale nRVCS, the pixel expansion is $(g-1)m$ , the computation complexity for reconstructing the secret image is $O(k-1)$ , which only usually perform $k-1$ OR operations to $k$ shares. The quality of reconstruction the secret image is $1/m(g-1)$ . Here, we compare our schemes with grayscale nRVCS in terms of reconstruction complexity, contrast, shares held by each participant, pixel expansion and variant aspect ratio. Next Table 6.1 is a comparison between grayscale nRGVCS and our proposed RVCSs for grayscale image.

Table 6.1 A comparison of properties among proposed $(k,n)$-RVCSs for grayscale image

| | Greyscale nRVCS | Section 3's scheme | Section 4's scheme | Section 5's scheme |
|---|---|---|---|---|
| Complexity | $O(k-1)$ | $O(k-1)$ | $O(k-1)$ | $O(k-1)$ |
| Shares held | 1 | $m$ | $m^2$ | 2 |

| Number of runs | 1 | $m$ | $m$ | 2 |
|---|---|---|---|---|
| Contrast | $1/m(g-1)$ | $1/g-1$ | $1/g-1$ | $1/g-1$ |
| Pixel expansions | $(g-1)m$ | $(g-1)$ | $(g-1)m$ | $(g-1)2^{k-1}\binom{n}{k}$ |
| Variant aspect ratio | $(g-1)m$ | $(g-1)$ | $(g-1)m$ | $(g-1)$ |

From Table 6.1 above, in traditional (*k*, *n*)-nRGVCS, the pixel expansion is $(g-1)m$, storage requirement is $(g-1)m$ for each participant, the quality (contrast) of reconstruction is $1/m(g-1)$. Each participant holds one share. In our proposed schemes, pixel expansion of three schemes is $(g-1)$, $(g-1)m$, and $(g-1)2^{k-1}\binom{n}{k}$, respectively. The contrast is $1/(g-1)$. The storage requirement of three schemes is $(g-1)m$, $(g-1)m^2$, and $(g-1)2^{k}\binom{n}{k}$, respectively. The number of shares held for each participant is $m$, $m$, and 2, respectively.

For easy lookup and comparison, the contrast of proposed schemes is higher than that of traditional GVCS with reversing. The storage requirement of our scheme in section 3 is equivalent to that of traditional GVCS with reversing. It is easy to verify that the value of storage requirement in section 4 is lower than that of the size of traditional GVCS with reversing when $k \geq \frac{n}{8}$, $k \geq 8$. Although the scheme in section 4 has large storage requirement, it can apply the case of basis matrices that are not perfect black.

### 6.2 *Comparison with Boolean-based secret sharing schemes*

Some secret sharing schemes in [15-20] only need one share for each participant and one run to obtain better contrast by Boolean-based reconstruction. These Boolean-based schemes can be divided to two types. One type is XOR-based nRVCS [15], the other is based on Boolean operation [16-20]. We will compare our proposed scheme with typical Boolean -based secret sharing schemes in Table 6.2.

Table 6.2 Comparison with Boolean -based secret sharing schemes for single pixel value

| | Image type | Complexity | Contrast | Storage requirement | Compatibility | Variant aspect ratio |
|---|---|---|---|---|---|---|
| Tuyls et al.[15] | binary | $O(k)$ | 1(only for (*n*, *n*)) | $m$ | No | $m$ |
| Wang et al. [16] | greyscale | $O(k)$ | 1(only for (*n*, *n*)) | 1 | No | 1 |
| Chao et al. [19] | greyscale | $O(k)$ | 1 | $\frac{2(n-k+1)}{n}$ a | No | 1 |
| Proposed RVCS | greyscale | $O(k)$ | 1 for computing; $1/(g-1)$ for stacking | $m(g-1)$ | Yes | 1 for computing; $g$-1 for stacking |

a. Extra shadows-assignment matrix H with $n \times \binom{n}{k-1}$ is needed.

According to the above comparisons, the advantages of our constructions can be seen as follows. Firstly, if the reconstruction is based on computing, our scheme can have ideal contrast, which is equal to the schemes in [19]. Furthermore, comparing with XOR-based secret sharing scheme, our schemes have the advantage that even if we do not have the copy machine with reversing operation, our schemes above can reconstruct the secret image by stacking the shares directly.

## 7 .Conclusions

We first use *within-block-column-permutation* method to design a greyscale visual cryptography scheme, which has the same pixel expansion and contrast as the existing GVCS. Using our greyscale nRVCS, we then propose three optimal grey levels RVCS schemes, which can satisfy different user requirement.

**Acknowledgments:** The authors thank Professor Xiaobo Li of University of Alberta for his valuable suggestions and help. This research was supported in part by the National Natural Science Foundation of China (Grant No. 61170032), and was also supported by the Testbed@TWISC, National Science Council under the Grant NSC 100-2219-E-006-001.

## References

[1] M. Naor and A. Shamir, Visual cryptography, in Process Advances in Cryptology (*EUROCRYPT'94*), 1995, vol. 950, pp. 1-12, Springer-Verlag, Lecture Notes in Computer Science.

[2] E. R. Verheul and H. C. A. Van Tilborg, Constructions and properties of k-out-of-n visual secret sharing schemes, Designs, Codes and Cryptography, vol.11, pp.179-196, 1997.

[3] C.Blundo and A. de. Santis, Visual cryptography schemes with perfect reconstruction of black pixels, Computer and Graphics, vol.22, no.4, pp.449-455, 1998.

[4] C. Blundo, A. De Bonis, and A. De Santis, Improved schemes for visual cryptography, Designs, Codes and Cryptograph, vol. 24, pp.255 -278, 2001.
[5] D.Q. Viet and K. Kurosawa, Almost ideal contrast visual cryptography with reversing, in Proceeding of Topics in Cryptology(*CT-RSA 2004*),2004, vol. 2964, pp. 353-365,Springer-Verlag, Lecture note in Computer Science.
[6] S. Cimato, A. De Santis, A.L. Ferrara ,and B. Masucci, Ideal contrast visual cryptography schemes with reversing, Information Processing Letters, vol. 93, no.4, pp.199-206,2005.
[7] C.M. Hu and W.G.Tzeng, Compatible Ideal Contrast Visual Cryptography Schemes with Reversing, Information Security, vol. 3650, pp.300-313, 2005.
[8] C.N. Yang, C.C. Wang ,and T.S. Chen, Real Perfect Contrast Visual Secret Schemes with Reversing, Applied Cryptography and Network Security 2006, 2006, vol. 3989, pp. 433-447,Springer-Verlag, Lecture Notes in Computer Science.
[9] C.N. Yang, C.C. Wang ,and T.S.Chen, Visual Cryptography Schemes with Reversing, The Computer Journal, vol. 51, no.6, pp.710-722, 2008.
[10] X.Q. Tan, Two kinds of ideal contrast visual cryptography schemes, 2009 International Conference on Signal Processing Systems ( *ICSPS 2009*), 2009, pp. 450-453, IEEE.
[11] H.B. Zhang, X.F. Wang ,and Y.P. Huang, A Novel Ideal Contrast Visual Secret Sharing Scheme with Reversing, Journal of Multimedia, vol.4, no.3,pp.104-111,2009.
[12] L.G. Fang, Y.M. Li, and B. Yu, Multi-Secret Visual Cryptography Based On Reversed Images, Proceedings of the Third International Conference on Information and Computing Science (*ICIC 2010*), 2010, pp.195-198, IEEE.
[13] I. Muecke, Greyscale and colour visual cryptography, Thesis of degree of master of computer science, Dalhouse uinversity-Daltech, Canada, 1999.
[14] C. Blundo, A. De Santis, and M. Naor, Visual cryptography for grey level images, Information Processing Letters, vol. 75, pp.255-259, 2000.
[15] P. Tuyls, H.D.L. Hollmann, J.H.van Lint, and L. Tolhuizen. Xor-based visual cryptography Schemes, Designs, Codes and Cryptography, vol.37, pp.169-186, 2005.
[16] D.S. Wang, L. Zhang, N. Ma, and X. Li, Two secret sharing schemes based on Boolean operations. Pattern Recognition, vol. 40, pp. 2776-2785, 2007.

[17] C.C.Chang, C.C. Lin, T.H.N. Le, and H.B. Le, A Probabilistic Visual Secret Sharing Scheme for Grayscale Images with Voting Strategy, Proceedings of the International Symposium on Electronic Commerce and Security, 2008, pp.184-188, IEEE.

[18] M. Ulutas, V.V. Nabiyev, and G. Ulutas, A PVSS scheme based on Boolean operations with improved contrast, 2009 International Conference on Network and Service Security , 2009, pp.1-5, IEEE.

[19] K. Y. Chao and J. C. Lin, Secret image sharing: a Boolean-operations-based approach combining benefits of polynomial-based and fast approaches, International Journal of Pattern Recognition and Artificial Intelligence,vol.23, no.2, pp.263-285, 2009.

**[20]** D. S. Wang and L. Dong, XOR-based visual cryptography (chapter 6), Visual cryptography and secret image sharing, edited by S. Cimato and C. N. Yang. CRC press. 2011.

## Appendix A. Comparison table, notation and its description

Table 1 A comparison of properties among typical ($k$, $n$)-RVCSs

| | | Viet and Kurosawa's scheme | Cimato et al.'s first scheme | Hu et al. 's scheme | Yang et al.' A method | Yang et al.' B method |
|---|---|---|---|---|---|---|
| Compatibility | | ∨ | ∨ | ∨ | ∨ | ∨ |
| Scheme type | | PBVCS | PBVCS | PBVCS | PBVCS | nPBVCS |
| Complexity of reconstructed secret image | Number of reversing operations | $rk-1$ | $mk-1$ | $4k$ | $k(m-h+1)-1$ | $k\times(m+1)$ |
| | Number of stacking operations | $r+1$ | $m+1$ | $4k-1$ | $m-h+2$ | $m+1$ |
| Shares held by each participant | | $r$ | $m$ | 2 | $m(m-h+1)$ | $m\times m$ |
| Number of runs to achieve perfect contrast | | $r\to 8$ | $m$ | 2 | $m-h+1$ | $m$ |
| Contrast | | $1-(\frac{m-h}{m})^c$ | 1 | 1 | 1 | 1 |
| Pixel expansions | | $rm$ | $m$ | $2^{k-1}\binom{n}{k}$ | $m$ | $m$ |
| Variant aspect ratio | | $m$ | 1 | $2^{k-1}$ | $m$ | $m$ |

Table 2 Notations between VCS and $(k,n)$-GVCS

| Type of scheme | Abbreviation | Description |
|---|---|---|
| Binary $(k,n)$-VCS | $B_0$ and $B_1$ | $n\times m$ basis matrices. |
| | $m$ | Pixel expansion. |
| | $m-l$ | Hamming weight of the stacking result of any $k$ out of $n$ rows from matrix in $C_1$. |
| | $m-h$ | Hamming weight of the stacking result of any $k$ out of $n$ rows from matrix in $C_0$. |
| | $\alpha$ | Relative difference (contrast), $\alpha=(h-l)/m$ |
| $g$ grey-levels $(k,n)$-GVCS | $G^q$ | $n\times m_g$ basic matrices of the $q$-*th* grey –levels, $q=0,\cdots,g-1$. |
| | $m_g$ | Pixel expansion. |
| | $\alpha^{(q+1,q)}$ | The relative difference (or contrast) $\alpha^{(q+1,q)}$ between $q+1$-*th* and $q$-*th* grey-levels, $q=0,\cdots,g-2$. |

Table 3 Notation and its description

| Abbreviation | Description |
|---|---|
| $\otimes$ | XOR operation. |
| $\oplus$ | OR operation. |
| $H(.)$ | The Hamming weight function. |
| $OR(B_i \mid t)$ | The "OR"-ed $t$ rows in $B_i$ $i$ =0, 1. |
| $OR(G^q \mid t)$ | The "OR"-ed $t$ rows in $G^q$ ( $01 \le q \le g-1$) . |
| $\circ$ | Concatenation operation. |
| $\overline{T}$ | Not operation ( or reversing operation) to $T$ . |
| $P$ | A original Pixel. |
| $\tilde{P}$ | The reconstructed pixel. |
| $P^q$ | The reconstructed pixel of $P$ in $G^q$ ( $0 \le q \le g-1$) . |
| $X$ | A random variable. |
| $r$ | Run times. |
| $E(X)$ | The expected value for $X$ . |
| $\Gamma(\cdot)$ | Cyclically shifts right function. |
| VCS | Visual cryptography scheme |
| RVCS | Reversing-based VCS( or VCS with reversing) . |
| nRVCS | Traditional non- Reversing-based VCS. |
| GVCS | Greyscale visual cryptography scheme. |
| PBVCS | Perfect black visual cryptography scheme . |
| nPBVCS | Non perfect black visual cryptography scheme. |
| nRGVCS | Greyscale visual cryptography scheme without reversion. |
| RGVCS | Reversing-based greyscale VCS. |
| WBCP | Within-block-column- permutation. |
| PBRVCS | Perfect black RVCS (reversing-based VCS) . |
| RGPBVCS | Reversing-based greyscale PBVCS. |
| RGnPBVCS | Reversing-based greyscale nPBVCS. |
| nR-WBCP -GVCS | Greyscale nRVCS within-block-column- permutation. |

# Appendix B. Experimental results of the section 3

## 1. Brief review of Cimato et al.' perfect black scheme with reversing

We briefly describe Cimato et al.'s perfect black RVCS (PBRVCS), the construction procedure of Cimato et al.'s scheme is given in Table B-1 as follows.

Table B-1 Distribution phase and reconstruction phase of Cimato et al.'s $(k, n)$ scheme

| **Distribution phase** | **Reconstruction phase** |
|---|---|
| **Step 1:** The dealer D randomly chooses a matrix $S^0 = [s_{i,j}]$ in $C^0$ ($S^1$ in $C^1$, resp.)<br>**Step 2:** For each participant $i$, consider the $m$ bits $s_{i,1}, s_{i,2}, \ldots, s_{i,m}$ composing the $i$ –th row of $S^0$ and $S^1$, for each $j = 1, \ldots, m$, put a white (black, resp.) pixel on the transparency $t_{i,j}$ if $s_{i,j} = 0$ ($s_{i,j} = 1$, resp.). | **Step 1:** Any $k$ participants in $Q$ reconstruct the secret image by computing: $T_j = OR(t_{i_1,j}, \ldots, t_{i_k,j})$, for $j = 1, \ldots, m$.<br>**Step 2:** $\tilde{P} = \overline{(OR(\overline{T}_1 \oplus \ldots \oplus \overline{T}_m))}$ ,which is the reconstructed secret image. |

***Example B-1 (continuation of Example 2.1):***

The basic matrix $B_0$ and $B_1$ of a (2, 3)-VCS are

$$B_0 = \begin{bmatrix} 1 & 1 & 0 \\ 1 & 1 & 0 \\ 1 & 1 & 0 \end{bmatrix}, B_1 = \begin{bmatrix} 0 & 1 & 1 \\ 1 & 0 & 1 \\ 1 & 1 & 0 \end{bmatrix}.$$

The collections $C_0$ and $C_1$ are obtained by permuting the columns of the corresponding basis matrix ($B_0$ for $C_0$, and $B_1$ for $C_1$) in all possible ways. A Cimato et al.'s (2, 3) scheme is shown in next Table B-2.

Table B-2 Distribution phase and reconstruction phase of a Cimato et al.'s (2, 3) scheme

| **Distribution phase** | **Reconstruction phase** |
|---|---|
| **White Pixel:**<br>For Participant 1, $\overbrace{t_{1,1} = 1}^{1st\ run}$, $\overbrace{t_{1,2} = 1}^{2nd\ run}$, $\overbrace{t_{1,3} = 0}^{3rd\ run}$<br>For Participant 2, $t_{2,1}$=1, $t_{2,2}$=1, $t_{2,3}$=0<br>For Participant 3, $t_{3,1}$=1, $t_{3,2}$=1, $t_{3,3}$=0 | **White Pixel:**<br>Participant 1 + participant 2,<br>$T_1 = OR(t_{1,1}, t_{2,1}) = 1, T_2 = OR(t_{1,2}, t_{2,2}) = 1$<br>$T_3 = OR(t_{1,3}, t_{2,3}) = 0, U = \overline{(OR(\bar{1}, \bar{1}, \bar{0}))} = 0$.<br>Participant 1 + participant 3,<br>$T_1 = OR(t_{1,1}, t_{3,1}) = 1, T_2 = OR(t_{1,2}, t_{3,2}) = 1$,<br>$T_3 = OR(t_{1,3}, t_{3,3}) = 0, U = \overline{(OR(\bar{1}, \bar{1}, \bar{0}))} = 0$.<br>Participant 2 + participant 3,<br>$T_1 = OR(t_{2,1}, t_{3,1}) = 1, T_2 = OR(t_{2,2}, t_{3,2}) = 1$,<br>$T_3 = OR(t_{2,3}, t_{3,3}) = 0, U = \overline{(OR(\bar{1}, \bar{1}, \bar{0}))} = 0$. |

| Black Pixel : | Black Pixel : |
| --- | --- |
| For Participant 1, $\underbrace{t_{1,1}=0}_{1st\ run}$, $\underbrace{t_{1,2}=1}_{2nd\ run}$, $\underbrace{t_{1,3}=1}_{3rd\ run}$<br>For Participant 2, $t_{2,1}$=1, $t_{2,2}$=0, $t_{2,3}$=1<br>For Participant 3, $t_{3,1}$=1, $t_{3,2}$=1 $t_{3,3}$=0 | Participant 1 + participant 2,<br>$T_1 = OR(t_{1,1}, t_{2,1}) = 1, T_2 = OR(t_{1,2}, t_{2,2}) = 1,$<br>$T_3 = OR(t_{1,3}, t_{2,3}) = 1, U = \overline{(OR(\bar{1}, \bar{1}, \bar{1}))} = 1.$<br>Participant 1 + participant 3,<br>$T_1 = OR(t_{1,1}, t_{3,1}) = 1, T_2 = OR(t_{1,2}, t_{3,2}) = 1,$<br>$T_3 = OR(t_{1,3}, t_{3,3}) = 1, U = \overline{(OR(\bar{1}, \bar{1}, \bar{1}))} = 1.$<br>Participant 2 + participant 3,<br>$T_1 = OR(t_{2,1}, t_{3,1}) = 1, T_2 = OR(t_{2,2}, t_{3,2}) = 1,$<br>$T_3 = OR(t_{2,3}, t_{3,3}) = 1, U = \overline{(OR(\bar{1}, \bar{1}, \bar{1}))} = 1.$ |

Table B-2 shows the whiteness percentage of the white secret pixel can be improved to full white (100%) within three runs. It is evident that whiteness percentage of the black secret pixels is still full zero (0%) because we use the PBVCS. So it is a really ideal contrast scheme when finishing 3 runs. Namely, $\alpha = \frac{h-l}{m} = 1$.

The reconstruction phase of the $(k, n)$-PBRVCS for Cimato et al.' method, operations of stacking any $k$ shares equal to $mk-1$ OR operations. Then $m+1$ NOT operations are required to finish $m$ runs. Each participant hold $m$ shares. The size of shares becomes $m$ times larger than that of the original secret image. The size of the reconstructed image is the same as that of the original image.

## 2. *The analysis of directly extending scheme above to RGVCS*

We will give an example of reversing-based three grey levels (2, 3)-GVCS by directly expending /using Cimato et al.'s scheme as follows.

***Example B-2 (continuation of Example 2.2):***

The basis matrices of a deterministic (2, 3)-GVCS with three grey-levels are $G^0, G^1$, and $G^2$.

$$G^0 = B_0 \circ B_0 = \begin{bmatrix} 1 & 1 & 0 \\ 1 & 1 & 0 \\ 1 & 1 & 0 \end{bmatrix} \circ \begin{bmatrix} 1 & 1 & 0 \\ 1 & 1 & 0 \\ 1 & 1 & 0 \end{bmatrix} = \begin{bmatrix} 1 & 1 & 0 & 1 & 1 & 0 \\ 1 & 1 & 0 & 1 & 1 & 0 \\ 1 & 1 & 0 & 1 & 1 & 0 \end{bmatrix}, \quad G^1 = B_0 \circ B_1 = \begin{bmatrix} 1 & 1 & 0 \\ 1 & 1 & 0 \\ 1 & 1 & 0 \end{bmatrix} \circ \begin{bmatrix} 1 & 1 & 0 \\ 1 & 0 & 1 \\ 0 & 1 & 1 \end{bmatrix} = \begin{bmatrix} 0 & 1 & 1 & 1 & 1 & 0 \\ 0 & 1 & 1 & 1 & 0 & 1 \\ 0 & 1 & 1 & 0 & 1 & 1 \end{bmatrix},$$

$$G^2 = B_1 \circ B_1 = \begin{bmatrix} 1 & 1 & 0 \\ 1 & 0 & 1 \\ 0 & 1 & 1 \end{bmatrix} \circ \begin{bmatrix} 1 & 1 & 0 \\ 1 & 0 & 1 \\ 0 & 1 & 1 \end{bmatrix} = \begin{bmatrix} 1 & 1 & 0 & 1 & 1 & 0 \\ 1 & 0 & 1 & 1 & 0 & 1 \\ 0 & 1 & 1 & 0 & 1 & 1 \end{bmatrix}$$

We directly extend Cimato et al.'s binary RVCS [6] to construct GVCS with reversing. Table B-3 shows distribution phase (reconstruction phase) of such scheme. Example B-5 shows the experimental result.

Table B-3 Distribution phase and reconstruction phase of a (2, 3)-GVCS with reversing

| Distribution phase | | Reconstruction phase |
|---|---|---|
| **Grey level 1:**<br>For Participant 1,<br>$\overbrace{t_{1,1}=1}^{1st\ run}, \overbrace{t_{1,2}=1}^{2nd\ run}, \overbrace{t_{1,3}=0}^{3rd\ run}, \overbrace{t_{1,4}=1}^{4th\ run}, \overbrace{t_{1,5}=1}^{5th\ run}, \overbrace{t_{1,6}=0}^{6th\ run}$<br>For Participant 2,<br>$t_{2,1}=1$, $t_{2,2}=1$, $t_{2,3}=0$, $t_{2,4}=1$, $t_{2,5}=1$, $t_{2,6}=0$<br>For Participant 3,<br>$t_{3,1}=1$, $t_{3,2}=1$, $t_{3,3}=0$, $t_{3,4}=1$, $t_{3,5}=1$, $t_{3,6}=0$ | | **Grey level 1:**<br>Participant 1 + participant 2,<br>$T_1 = OR(t_{1,1}, t_{2,1}) = 1$, $T_2 = 1$, $T_3 = 0$, $T_4 = 1$, $T_5 = 1$, $T_6 = 0$, $U = \overline{(OR(\bar{1},\bar{1},\bar{0},\bar{1},\bar{1},\bar{0}))} = 0$.<br>Participant 1 + participant 3,<br>$T_1 = OR(t_{1,1}, t_{3,1}) = 1$, $T_2 = 1$, $T_3 = 0$, $T_4 = 1$, $T_5 = 1$, $T_6 = 0$, $U = \overline{(OR(\bar{1},\bar{1},\bar{0},\bar{1},\bar{1},\bar{0}))} = 0$.<br>Participant 2 + participant 3,<br>$T_1 = OR(t_{2,1}, t_{3,1}) = 1$, $T_2 = 1$, $T_3 = 0$, $T_4 = 1$, $T_5 = 1$, $T_6 = 0$, $U = \overline{(OR(\bar{1},\bar{1},\bar{0},\bar{1},\bar{1},\bar{0}))} = 0$. |
| **Grey level 2:**<br>For Participant 1,<br>$\overbrace{t_{1,1}=1}^{1st\ run}, \overbrace{t_{1,2}=1}^{2nd\ run}, \overbrace{t_{1,3}=0}^{3rd\ run}, \overbrace{t_{1,4}=1}^{4th\ run}, \overbrace{t_{1,5}=1}^{5th\ run}, \overbrace{t_{1,6}=0}^{6th\ run}$<br>For Participant 2,<br>$t_{2,1}=1$, $t_{2,2}=1$, $t_{2,3}=0$, $t_{2,4}=1$, $t_{2,5}=0$, $t_{2,6}=1$<br>For Participant 3,<br>$t_{3,1}=1$, $t_{3,2}=1$, $t_{3,3}=0$, $t_{3,4}=0$, $t_{3,5}=1$, $t_{3,6}=1$ | | **Grey level 2:**<br>Participant 1 + participant 2,<br>$T_1 = OR(t_{1,1}, t_{2,1}) = 1$, $T_2 = 1$, $T_3 = 0$, $T_4 = 1$, $T_5 = 1$, $T_6 = 1$, $U = \overline{(OR(\bar{1},\bar{1},\bar{0},\bar{1},\bar{1},\bar{1}))} = 0$.<br>Participant 1 + participant 3,<br>$T_1 = OR(t_{1,1}, t_{3,1}) = 1$, $T_2 = 1$, $T_3 = 0$, $T_4 = 1$, $T_5 = 1$, $T_6 = 1$, $U = \overline{(OR(\bar{1},\bar{1},\bar{0},\bar{1},\bar{1},\bar{1}))} = 0$.<br>Participant 2 + participant 3,<br>$T_1 = OR(t_{2,1}, t_{3,1}) = 1$, $T_2 = 1$, $T_3 = 0$, $T_4 = 1$, $T_5 = 1$, $T_6 = 1$, $U = \overline{(OR(\bar{1},\bar{1},\bar{0},\bar{1},\bar{1},\bar{1}))} = 0$. |
| **Grey level 3:**<br>For Participant 1,<br>$\overbrace{t_{1,1}=1}^{1st\ run}, \overbrace{t_{1,2}=1}^{2nd\ run}, \overbrace{t_{1,3}=0}^{3rd\ run}, \overbrace{t_{1,4}=1}^{4th\ run}, \overbrace{t_{1,5}=1}^{5th\ run}, \overbrace{t_{1,6}=0}^{6th\ run}$<br>For Participant 2,<br>$t_{2,1}=1$, $t_{2,2}=0$, $t_{2,3}=1$, $t_{2,4}=1$, $t_{2,5}=0$, $t_{2,6}=1$<br>For Participant 3,<br>$t_{3,1}=0$, $t_{3,2}=1$, $t_{3,3}=1$, $t_{3,4}=0$, $t_{3,5}=1$, $t_{3,6}=1$ | | **Grey level 3:**<br>Participant 1 + participant 2,<br>$T_1 = OR(t_{1,1}, t_{2,1}) = 1$, $T_2 = 1$, $T_3 = 1$, $T_4 = 1$, $T_5 = 1$, $T_6 = 1$, $U = \overline{(OR(\bar{1},\bar{1},\bar{1},\bar{1},\bar{1},\bar{1}))} = 1$.<br>Participant 1 + participant 3,<br>$T_1 = OR(t_{1,1}, t_{3,1}) = 1$, $T_2 = 1$, $T_3 = 1$, $T_4 = 1$, $T_5 = 1$, $T_6 = 1$, $U = \overline{(OR(\bar{1},\bar{1},\bar{1},\bar{1},\bar{1},\bar{1}))} = 1$.<br>Participant 2 + participant 3,<br>$T_1 = OR(t_{2,1}, t_{3,1}) = 1$, $T_2 = 1$, $T_3 = 1$, $T_4 = 1$, $T_5 = 1$, $T_6 = 1$, $U = \overline{(OR(\bar{1},\bar{1},\bar{1},\bar{1},\bar{1},\bar{1}))} = 1$. |

From Table B-3 we can see that pixel with grey level 1 and pixel with grey level 2 are both reconstructed as white pixel. The original secret image cannot be correctly reconstructed. This means directly using Cimato et al.'s binary scheme with reversing to perform three grey levels (2, 3)-GVCS with reversing is failed.

***Example B-3 (continuation of Example 2.2) :***

***Permutation method II:*** $\overleftrightarrow{G}^0 = \overleftrightarrow{B_0} \circ \overleftrightarrow{B_0}$ **,** $\overleftrightarrow{G}^1 = \overleftrightarrow{B_0} \circ \overleftrightarrow{B_1}$ **,** $\overleftrightarrow{G}^2 = \overleftrightarrow{B_1} \circ \overleftrightarrow{B_1}$ **.** When distributing the pixel in each share, we choose two pixels for each share, each pixel comes from the same place in each component.

Table B-4 The distribution phase under Permutation method II

| Grey levels | Chosen matrices | 1st run | 2nd run | 3rd run |
|---|---|---|---|---|
| 1 | $\left[\begin{array}{ccc\|ccc} 0 & 1 & 1 & 0 & 1 & 1 \\ 0 & 1 & 1 & 0 & 1 & 1 \\ 0 & 1 & 1 & 0 & 1 & 1 \end{array}\right]$ | $t_{1,1}$ =00 | $t_{1,2}$ =11 | $t_{1,3}$ =11 |
| | | $t_{2,1}$ =00 | $t_{2,2}$ =11 | $t_{2,3}$ =11 |
| | | $t_{3,1}$ =00 | $t_{3,2}$ =11 | $t_{3,3}$ =11 |
| 2 | $\left[\begin{array}{ccc\|ccc} 0 & 1 & 1 & 0 & 1 & 1 \\ 0 & 1 & 1 & 1 & 0 & 1 \\ 0 & 1 & 1 & 1 & 1 & 0 \end{array}\right]$ | $t_{1,1}$ =00 | $t_{1,2}$ =11 | $t_{1,3}$ =11 |
| | | $t_{2,1}$ =01 | $t_{2,2}$ =10 | $t_{2,3}$ =11 |
| | | $t_{3,1}$ =01 | $t_{3,2}$ =11 | $t_{3,3}$ =10 |
| 3 | $\left[\begin{array}{ccc\|ccc} 0 & 1 & 1 & 0 & 1 & 1 \\ 1 & 0 & 1 & 1 & 0 & 1 \\ 1 & 1 & 0 & 1 & 1 & 0 \end{array}\right]$ | $t_{1,1}$ =00 | $t_{1,2}$ =11 | $t_{1,3}$ =11 |
| | | $t_{2,1}$ =11 | $t_{2,2}$ =00 | $t_{2,3}$ =11 |
| | | $t_{3,1}$ =11 | $t_{3,2}$ =11 | $t_{3,3}$ =00 |

Table B-5 The reconstruction phase under Permutation method II

Participant 1 + participant 2

| Grey levels | $T_1 = OR(t_{1,1}, t_{2,1})$ | $T_2 = OR(t_{1,2}, t_{2,2})$ | $T_3 = OR(t_{1,3}, t_{2,3})$ | $\tilde{P} = \overline{(OR(\overline{T_1}, \overline{T_2}, \overline{T_3}))}$ |
|---|---|---|---|---|
| 1 | 00 | 11 | 11 | 00 |
| 2 | 01 | 11 | 11 | 01 |
| 3 | 11 | 11 | 11 | 11 |

Participant 1 + participant 3

| Grey levels | $T_1 = OR(t_{1,1}, t_{3,1})$ | $T_2 = OR(t_{1,2}, t_{3,2})$ | $T_3 = OR(t_{1,3}, t_{3,3})$ | $\tilde{P} = \overline{(OR(\overline{T_1}, \overline{T_2}, \overline{T_3}))}$ |
|---|---|---|---|---|
| 1 | 00 | 11 | 11 | 00 |
| 2 | 01 | 11 | 11 | 01 |
| 3 | 11 | 11 | 11 | 11 |

Participant 2 + participant 3

| Grey levels | $T_1 = OR(t_{2,1}, t_{3,1})$ | $T_2 = OR(t_{2,2}, t_{3,2})$ | $T_3 = OR(t_{2,3}, t_{3,3})$ | $\tilde{P} = \overline{(OR(\overline{T_1}, \overline{T_2}, \overline{T_3}))}$ |
|---|---|---|---|---|
| 1 | 00 | 11 | 11 | 00 |
| 2 | 01 | 11 | 11 | 01 |
| 3 | 11 | 11 | 11 | 11 |

***Example B-4 (continuation of Example 2.2):***

***Permutation method III:*** $\overleftrightarrow{G}^0 = \overleftrightarrow{B}_0 \circ \overleftrightarrow{B}_0$, $\overleftrightarrow{G}^1 = \overleftrightarrow{B}_0 \circ \overleftrightarrow{B}_1$, $\overleftrightarrow{G}^2 = \overleftrightarrow{B}_1 \circ \overleftrightarrow{B}_1$. When distributing the pixel in each share, we choose two pixels for each share, each pixel is randomly chosen from each component.

Table B-6 The distribution phase under Permutation method III

| Grey levels | Chosen matrices | 1st run | 2nd run | 3rd run |
|---|---|---|---|---|
| 1 | $\left[\begin{array}{ccc\|ccc} 0&1&1&1&0&1 \\ 0&1&1&1&0&1 \\ 0&1&1&1&0&1 \end{array}\right]$ | $t_{1,1}$=01<br>$t_{2,1}$=01<br>$t_{3,1}$=01 | $t_{1,2}$=10<br>$t_{2,2}$=10<br>$t_{3,2}$=10 | $t_{1,3}$=11<br>$t_{2,3}$=11<br>$t_{3,3}$=11 |
| 2 | $\left[\begin{array}{ccc\|ccc} 0&1&1&1&0&1 \\ 0&1&1&0&1&1 \\ 0&1&1&1&1&0 \end{array}\right]$ | $t_{1,1}$=01<br>$t_{2,1}$=00<br>$t_{3,1}$=01 | $t_{1,2}$=10<br>$t_{2,2}$=11<br>$t_{3,2}$=11 | $t_{1,3}$=11<br>$t_{2,3}$=11<br>$t_{3,3}$=10 |
| 3 | $\left[\begin{array}{ccc\|ccc} 0&1&1&1&1&0 \\ 1&0&1&1&0&1 \\ 1&1&0&0&1&1 \end{array}\right]$ | $t_{1,1}$=01<br>$t_{2,1}$=11<br>$t_{3,1}$=10 | $t_{1,2}$=11<br>$t_{2,2}$=00<br>$t_{3,2}$=11 | $t_{1,3}$=10<br>$t_{2,3}$=11<br>$t_{3,3}$=01 |

Table B-7 The reconstruction phase under Permutation method III

| Participant 1 + participant 2 | | | | |
|---|---|---|---|---|
| Grey levels | $T_1 = OR(t_{1,1}, t_{2,1})$ | $T_2 = OR(t_{1,2}, t_{2,2})$ | $T_3 = OR(t_{1,3}, t_{2,3})$ | $\tilde{P} = \overline{(OR(\overline{T_1}, \overline{T_2}, \overline{T_3}))}$ |
| 1 | 01 | 10 | 11 | 00 |
| 2 | 01 | 11 | 11 | 01 |
| 3 | 11 | 11 | 11 | 11 |
| Participant 1 + participant 3 | | | | |
| Grey levels | $T_1 = OR(t_{1,1}, t_{3,1})$ | $T_2 = OR(t_{1,2}, t_{3,2})$ | $T_3 = OR(t_{1,3}, t_{3,3})$ | $\tilde{P} = \overline{(OR(\overline{T_1}, \overline{T_2}, \overline{T_3}))}$ |
| 1 | 01 | 10 | 11 | 00 |
| 2 | 01 | 11 | 11 | 01 |
| 3 | 11 | 11 | 11 | 11 |
| Participant 2 + participant 3 | | | | |
| Grey levels | $T_1 = OR(t_{2,1}, t_{3,1})$ | $T_2 = OR(t_{2,2}, t_{3,2})$ | $T_3 = OR(t_{2,3}, t_{3,3})$ | $\tilde{P} = \overline{(OR(\overline{T_1}, \overline{T_2}, \overline{T_3}))}$ |
| 1 | 00 | 11 | 11 | 00 |
| 2 | 01 | 11 | 11 | 01 |
| 3 | 11 | 11 | 11 | 11 |

***Example B-5:***

Experimental result of different permutation methods

| | Permutation Method I | Permutation Method II | Permutation Method III |
|---|---|---|---|
| Stacking result by participant 1 and 2 | 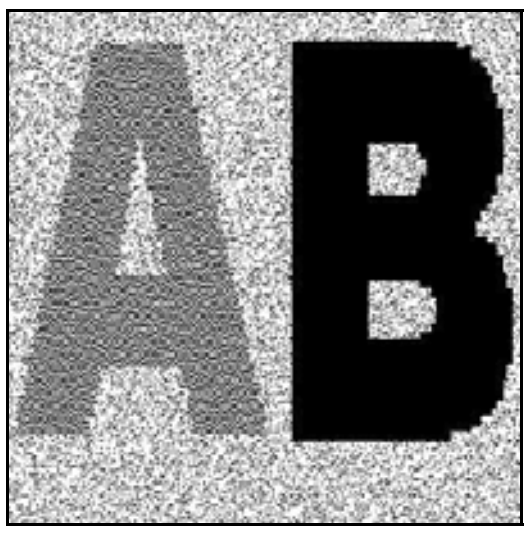 | 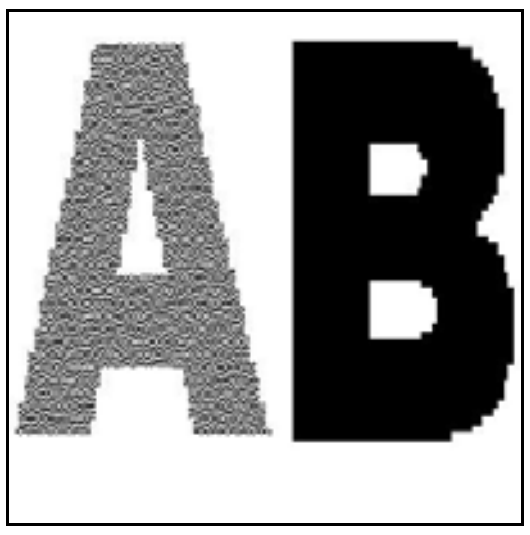 | 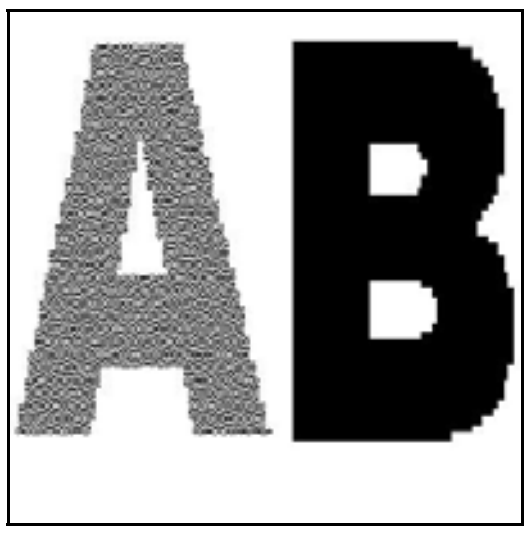 |
| $t_{1,1}$ held by participant 1 | 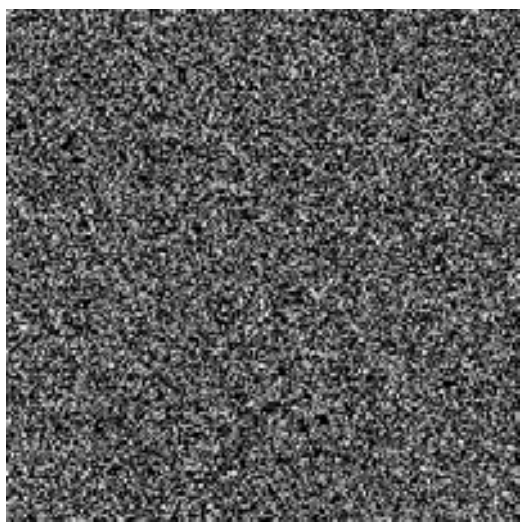 | 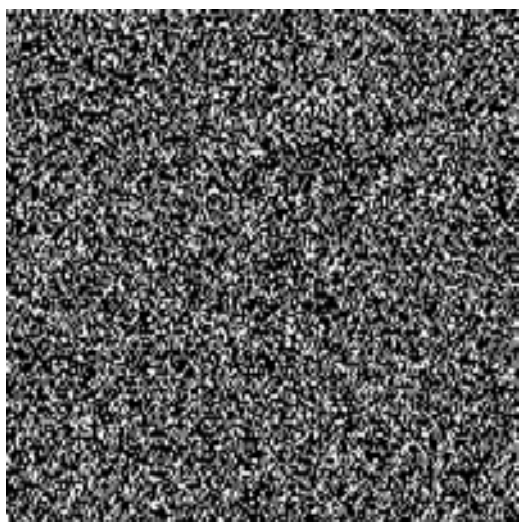 | 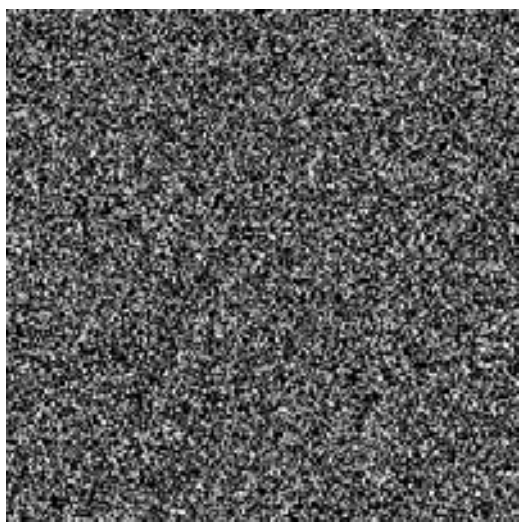 |
| $t_{1,2}$ held by participant 1 | 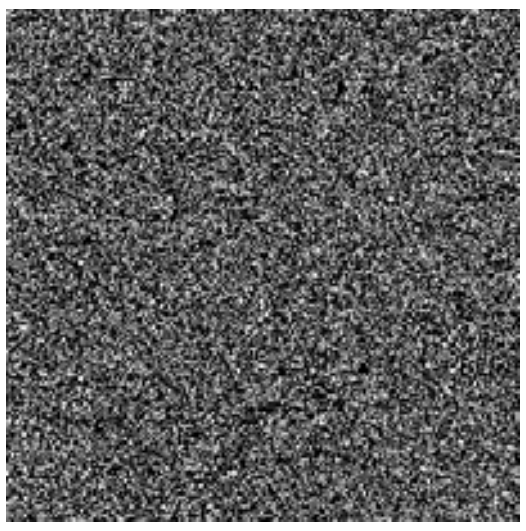 | 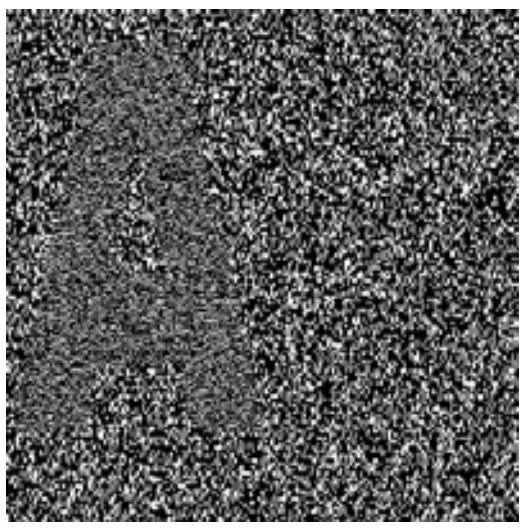 | 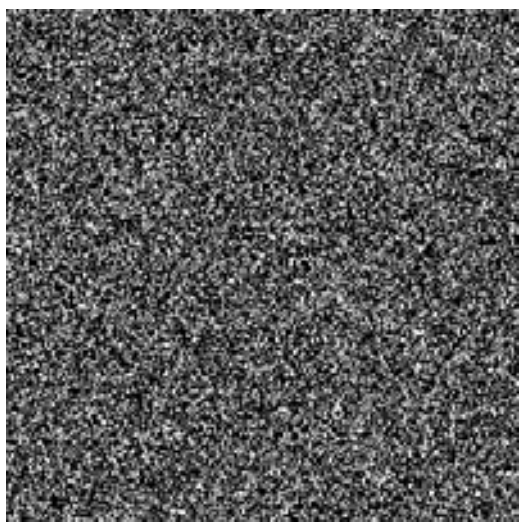 |
| $t_{1,3}$ held by participant 1 | 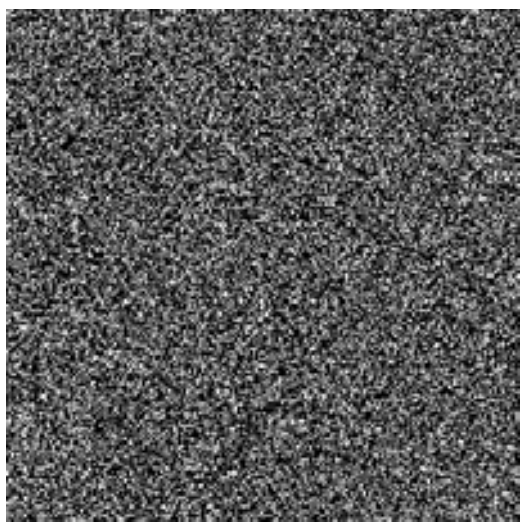 | 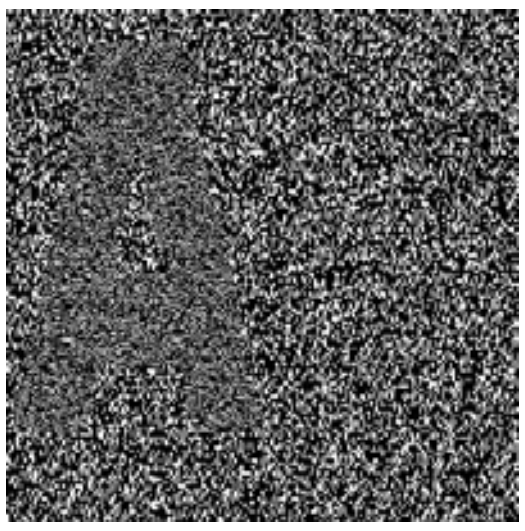 | 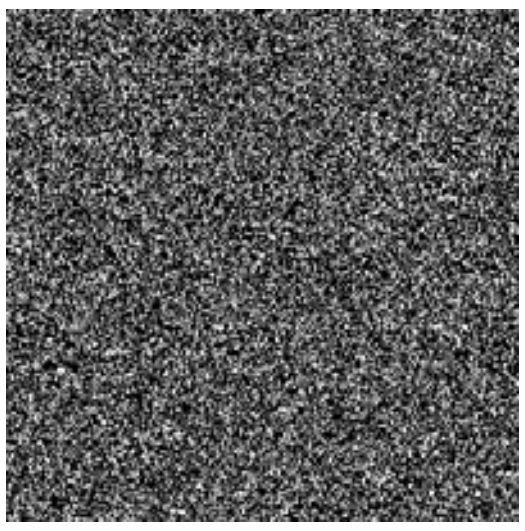 |

From the experimental result we can see that both methods can reconstruct the secret image correctly, but some secret information leaks out in Permutation Method II.

The reason is that in Permutation Method II, two components are permuted simultaneously. Since Grey level 1 and grey level 3 are composed by two same components, shares corresponding to grey level 1 and grey level 3 are 00 or 11. While grey level 2 is composed by two different components, shares of certain two participants will have 10 or 01. That is to say, there is difference between pixels corresponding to grey level 2 and pixels corresponding to grey level 1 and grey level 3, thus some information of grey level 2 leaks out.

# Appendix C. Analysis and experimental results of the section 4

In this appendix , we first briefly review Yang et al.'s NPBRVCS, and then analyze that we cannot get an optimal contrast when directly extending to grayscale RVCS.

Let the shadow image $S=[s_{ij_k}]$ , and the element $s_{ij\Psi}$ is the secret pixel $s_{ij}$ in a ($W\times H$)-pixel secret replaced by $m$ sub-pixels $(s_{i,j_1}\cdots s_{i,j_m})$, where $i\in[1,W], j\in[1,H]$ and $\Psi\in[1,m]$. We use symbol $\Gamma(\cdot)$ to represent cyclically shifts right one pixel in every $m$ sub-pixels (for a secret pixel) in the shadow image. The cyclic-shift operation is $\Gamma(S)=\lfloor\gamma(s_{i,j_k})\rfloor$,where $\gamma(\cdot)$ is a 1-bit cyclical right shift function. i.e. $\gamma(s_{ij_1}...s_{ij_m})=(s_{ij_m}s_{ij_1},...s_{ij_{m-1}})$ .

## 1. The Scheme of Ching-Nung Yang et al.

The distribution phase and reconstruction phase of Yang et al's scheme is given in Table C-1 as follows.

Table C-1 Distribution phase and reconstruction phase

| Distribution phase | | Reconstruction phase |
|---|---|---|
| To share a white (black, resp.) pixel, while $l\neq 0$ ,the dealer,<br>**Step1:** Given a secret image, the dealer performs an ($n,k$)-NPBVCS to generate $n$ shadows, $A_1^1,\dots,A_n^1$ for the first run.<br>**Step2:** The dealer generates the shadows $A_j^r=\Gamma(A_j^{r-1})$ for the $r$-th run, $r\in[2,m]$.Note that the shadow should be labeled as to which run it is, for easy management by the participant.<br>**Step3:** The dealer distributes $m$ shadows $A_j^1,\dots,A_j^m$ to Participant $j$ , $j\in[1,n]$ .<br>**Step4:** Finally, every participant holds $m$ shadows. | | To recover the secret within $m$ runs, at least $k$ participants, Participants $j_1,\dots,j_k$ , offer their $(k\times m)$ shadows $A_{j_1}^r,\dots,A_{j_k}^r$ , $r\in[1,m]$ , for reconstruction.<br>**Step1:** Stack the shadows $A_{j_1}^r,\dots,A_{j_k}^r$ to reconstruct the image $T_r$ in the $r$ th run.<br>**Step2:** Finish $m$ runs by using XOR operation to reconstruct $U'=T_1\otimes\cdots\otimes T_m$ .<br>**Step3:** If '$m-h$' is even (i.e. '$m-h$' is odd) then the reconstructed image is $\tilde{P}=U'$ ; otherwise the reconstructed image is $\tilde{P}=\overline{U'}$ . |

Distribution phase and reconstruction phase above is demonstrated as follows through an example (2, 3)-NPBVCS .

***Example C-1：Yang et al.'s (2, 3)-NPBVCS:***

The basis matrices are:

$$B_0 = \begin{bmatrix} 100 \\ 100 \\ 100 \end{bmatrix}, \quad B_1 = \begin{bmatrix} 100 \\ 010 \\ 001 \end{bmatrix}.$$

Table C-2 The distribution phase of a Yang et al. 's (2, 3)-NPBVCS above

| Pixel | 1st run | 2nd run | 3rd run |
|---|---|---|---|
| White | $t_{1,1}$=100 | $t_{1,2}=\Gamma(t_{1,1})$=010 | $t_{1,3}=\Gamma(t_{1,2})$=001 |
| | $t_{2,1}$=100 | $t_{2,2}=\Gamma(t_{2,1})$=010 | $t_{2,3}=\Gamma(t_{2,2})$=001 |
| | $t_{3,1}$=100 | $t_{3,2}=\Gamma(t_{3,1})$=010 | $t_{3,3}=\Gamma(t_{3,2})$=001 |
| Black | $t_{1,1}$=100 | $t_{1,2}=\Gamma(t_{1,1})$=010 | $t_{1,3}=\Gamma(t_{1,2})$=001 |
| | $t_{2,1}$=010 | $t_{2,2}=\Gamma(t_{2,1})$=001 | $t_{2,3}=\Gamma(t_{2,2})$=100 |
| | $t_{3,1}$=001 | $t_{3,2}=\Gamma(t_{3,1})$=100 | $t_{3,3}=\Gamma(t_{3,2})$=010 |

Table C-2 The Reconstruction phase of a Yang et al. 's (2, 3)-NPBVCS above

| Participant 1 + participant 2 | | | | |
|---|---|---|---|---|
| Pixel | $T_1=OR(t_{1,1},t_{2,1})$ | $T_2=OR(t_{1,2},t_{2,2})$ | $T_3=OR(t_{1,3},t_{2,3})$ | $\tilde{P}==\overline{T_1\otimes T_2\otimes T_3}$ |
| White | 100 | 010 | 001 | 000 |
| Black | 110 | 011 | 101 | 111 |
| **Participant 1 + participant 3** | | | | |
| Pixel | $T_1=OR(t_{1,1},t_{3,1})$ | $T_2=OR(t_{1,2},t_{3,2})$ | $T_3=OR(t_{1,3},t_{3,3})$ | $\tilde{P}==\overline{T_1\otimes T_2\otimes T_3}$ |
| White | 100 | 010 | 001 | 000 |
| Black | 101 | 110 | 011 | 111 |
| **Participant 2 + participant 3** | | | | |
| Pixel | $T_1=OR(t_{2,1},t_{3,1})$ | $T_2=OR(t_{2,2},t_{3,2})$ | $T_3=OR(t_{2,3},t_{3,3})$ | $\tilde{P}==\overline{T_1\otimes T_2\otimes T_3}$ |
| White | 100 | 010 | 001 | 000 |
| Black | 011 | 101 | 110 | 111 |

From the reconstruction phase we can achieve the perfect contrast when finishing two ($m$-$h$+1=2) runs.

## 2. *The analysis of directly extending scheme above to RGVCS*

We directly extend Yang et al.'s (2, 3)-NPBVCS to GVCS by the following example.

***Example C-2***:

The basis matrices for three grey levels (2, 3)-GVCS are:

$$G^0=\begin{bmatrix}1&0&0&1&0&0\\1&0&0&1&0&0\\1&0&0&1&0&0\end{bmatrix},\ G^1=\begin{bmatrix}1&0&0&1&0&0\\1&0&0&0&1&0\\1&0&0&0&0&1\end{bmatrix},\ G^2=\begin{bmatrix}1&0&0&1&0&0\\0&1&0&0&1&0\\0&0&1&0&0&1\end{bmatrix}.$$

The contrast is $\alpha_0=\frac{3-2}{6}=\frac{1}{6}, \alpha_1=\frac{4-3}{6}=\frac{1}{6}$, thus $\alpha_0=\alpha_1$.

When directly extending Yang et al.'s method to GVCS, $r=6$. Since $h=4$, $(m-h)=2$ is even, thus the reconstructed image is $U'$.

Table C-3 The distribution phase of a (2, 3)-GVCS

| Grey level | 1st run | 2nd run $t_{j,2}=\Gamma(t_{j,1})$ $j$=1,2,3 | 3rd run $t_{j,3}=\Gamma(t_{j,2})$ $j$=1,2,3 | 4th run $t_{j,4}=\Gamma(t_{j,3})$ $j$=1,2,3 | 5th run $t_{j,5}=\Gamma(t_{j,4})$ $j$=1,2,3 | 6th run $t_{j,6}=\Gamma(t_{j,5})$ $j$=1,2,3 |
|---|---|---|---|---|---|---|
| 0 | $t_{1,1}$=100100 | $t_{1,2}$=010010 | $t_{1,3}$=001001 | $t_{1,4}$=100100 | $t_{1,5}$=010010 | $t_{1,6}$=001001 |
| | $t_{2,1}$=100100 | $t_{2,2}$=010010 | $t_{2,3}$=001001 | $t_{2,4}$=100100 | $t_{2,5}$=010010 | $t_{2,6}$=001001 |
| | $t_{3,1}$=100100 | $t_{3,2}$=010010 | $t_{3,3}$=001001 | $t_{3,4}$=100100 | $t_{3,5}$=010010 | $t_{3,6}$=001001 |
| 1 | $t_{1,1}$=100100 | $t_{1,2}$=010010 | $t_{1,3}$=001001 | $t_{1,4}$=100100 | $t_{1,5}$=010010 | $t_{1,6}$=001001 |
| | $t_{2,1}$=100010 | $t_{2,2}$=010001 | $t_{2,3}$=101000 | $t_{2,4}$=010100 | $t_{2,5}$=001010 | $t_{2,6}$=000101 |
| | $t_{3,1}$=100001 | $t_{3,2}$=110000 | $t_{3,3}$=011000 | $t_{3,4}$=001100 | $t_{3,5}$=000110 | $t_{3,6}$=000011 |
| 2 | $t_{1,1}$=100100 | $t_{1,2}$=010010 | $t_{1,3}$=001001 | $t_{1,4}$=100100 | $t_{1,5}$=010010 | $t_{1,6}$=001001 |
| | $t_{2,1}$=010010 | $t_{2,2}$=001001 | $t_{2,3}$=100100 | $t_{2,4}$=010010 | $t_{2,5}$=001001 | $t_{2,6}$=100100 |
| | $t_{3,1}$=001001 | $t_{3,2}$=100100 | $t_{3,3}$=010010 | $t_{3,4}$=001001 | $t_{3,5}$=100100 | $t_{3,6}$=010010 |

Table C-4 The Reconstruction phase of a (2, 3)-GVCS

Participant 1 + participant 2

| Grey level | $T_1=OR(t_{1,1},t_{2,1})$ | $T_2=OR(t_{1,2},t_{2,2})$ | $T_3=OR(t_{1,3},t_{2,3})$ | $T_4=OR(t_{1,4},t_{2,4})$ | $T_5=OR(t_{1,5},t_{2,5})$ | $T_6=OR(t_{1,6},t_{2,6})$ | $\tilde{P}=T_1\otimes\cdots\otimes T_6$ |
|---|---|---|---|---|---|---|---|
| 0 | 100100 | 010010 | 001001 | 100100 | 010010 | 001001 | 000000 |
| 1 | 100110 | 010011 | 101001 | 110100 | 011010 | 001101 | 111111 |
| 2 | 110110 | 011011 | 101101 | 110110 | 011011 | 101101 | 000000 |

Participant 1 + participant 3

| Grey level | $T_1=OR(t_{1,1},t_{3,1})$ | $T_2=OR(t_{1,2},t_{3,2})$ | $T_3=OR(t_{1,3},t_{3,3})$ | $T_4=OR(t_{1,4},t_{3,4})$ | $T_5=OR(t_{1,5},t_{3,5})$ | $T_6=OR(t_{1,6},t_{3,6})$ | $\tilde{P}=T_1\otimes\cdots\otimes T_6$ |
|---|---|---|---|---|---|---|---|
| 0 | 100100 | 010010 | 001001 | 100100 | 010010 | 001001 | 000000 |
| 1 | 100101 | 110010 | 011001 | 101100 | 010110 | 001011 | 111111 |
| 2 | 101101 | 110110 | 011011 | 101101 | 110110 | 011011 | 000000 |

Participant 2 + participant 3

| Grey level | $T_1 = OR(t_{2,1},t_{3,1})$ | $T_2 = OR(t_{2,2},t_{3,2})$ | $T_3 = OR(t_{2,3},t_{3,3})$ | $T_4 = OR(t_{2,4},t_{3,4})$ | $T_5 = OR(t_{2,5},t_{3,5})$ | $T_6 = OR(t_{2,6},t_{3,6})$ | $\tilde{P} = T_1 \otimes \cdots \otimes T_6$ |
|---|---|---|---|---|---|---|---|
| 0 | 100100 | 010010 | 001001 | 100100 | 010010 | 001001 | 000000 |
| 1 | 100011 | 110001 | 111000 | 011100 | 001110 | 000111 | 111111 |
| 2 | 011011 | 101101 | 110110 | 011011 | 101101 | 110110 | 000000 |

From the reconstruction phase we can see that grey level 0 and grey level 2 cannot be distinguished, thus the method above is not used.

***Example C-3:***

The proposed three grey levels (2, 3)-GVCS using Yang et al.'s (2, 3)-NPBVCS. The basis matrices are

$$\overleftrightarrow{G}^0 = \begin{bmatrix} \overleftrightarrow{100} & \overleftrightarrow{100} \\ 100 & 100 \\ 100 & 100 \end{bmatrix}, \overleftrightarrow{G}^1 = \begin{bmatrix} \overleftrightarrow{100} & \overleftrightarrow{100} \\ 100 & 010 \\ 100 & 001 \end{bmatrix}, \overleftrightarrow{G}^2 = \begin{bmatrix} \overleftrightarrow{100} & \overleftrightarrow{100} \\ 010 & 010 \\ 001 & 001 \end{bmatrix}.$$

Since $(m-h)=1$, the reconstructed image is $\tilde{P}$.

Table C-5 The distribution phase of a (2, 3)-GVCS

| Grey level | 1st run | 2nd run | 3rd run |
|---|---|---|---|
| 0 | $t_{1,1}$=100100 | $t_{1,2}=\Gamma(100)\,\|\,\Gamma(100)$=010010 | $t_{1,3}=\Gamma(010)\,\|\,\Gamma(010)$=001001 |
| | $t_{2,1}$=100100 | $t_{2,2}=\Gamma(100)\,\|\,\Gamma(100)$=010010 | $t_{2,3}=\Gamma(010)\,\|\,\Gamma(010)$=001001 |
| | $t_{3,1}$=100100 | $t_{3,2}=\Gamma(100)\,\|\,\Gamma(100)$=010010 | $t_{3,3}=\Gamma(010)\,\|\,\Gamma(010)$=001001 |
| 1 | $t_{1,1}$=100100 | $t_{1,2}=\Gamma(100)\,\|\,\Gamma(100)$=010010 | $t_{1,3}=\Gamma(010)\,\|\,\Gamma(010)$=001001 |
| | $t_{2,1}$=100010 | $t_{2,2}=\Gamma(100)\,\|\,\Gamma(010)$=010001 | $t_{2,3}=\Gamma(010)\,\|\,\Gamma(001)$=001100 |
| | $t_{3,1}$=100001 | $t_{3,2}=\Gamma(100)\,\|\,\Gamma(001)$=010100 | $t_{3,3}=\Gamma(010)\,\|\,\Gamma(100)$=001010 |
| 2 | $t_{1,1}$=100100 | $t_{1,2}=\Gamma(100)\,\|\,\Gamma(100)$=010010 | $t_{1,3}=\Gamma(010)\,\|\,\Gamma(010)$=001001 |
| | $t_{2,1}$=010010 | $t_{2,2}=\Gamma(010)\,\|\,\Gamma(010)$=001001 | $t_{2,3}=\Gamma(001)\,\|\,\Gamma(001)$=100100 |
| | $t_{3,1}$=001001 | $t_{3,2}=\Gamma(001)\,\|\,\Gamma(001)$=100100 | $t_{3,3}=\Gamma(100)\,\|\,\Gamma(100)$=010010 |

Table C-6 The Reconstruction phase of a (2, 3)-GVCS

Participant 1 + participant 2

| Grey level | $T_1 = OR(t_{1,1},t_{2,1})$ | $T_2 = OR(t_{1,2},t_{2,2})$ | $T_3 = OR(t_{1,3},t_{2,3})$ | $\tilde{P} = \overline{T_1 \oplus T_2 \oplus T_3}$ |
|---|---|---|---|---|
| 0 | 100100 | 010010 | 001001 | 000000 |
| 1 | 100110 | 010011 | 001101 | 000111 |
| 2 | 110110 | 011011 | 101101 | 111111 |

Participant 1 + participant 3

| Grey level | $T_1 = OR(t_{1,1}, t_{3,1})$ | $T_2 = OR(t_{1,2}, t_{3,2})$ | $T_3 = OR(t_{1,3}, t_{3,3})$ | $\tilde{P} = \overline{T_1 \otimes T_2 \otimes T_3}$ |
|---|---|---|---|---|
| 0 | 100100 | 010010 | 001001 | 000000 |
| 1 | 100101 | 010110 | 001011 | 000111 |
| 2 | 101101 | 110110 | 011011 | 111111 |

Participant 2 + participant 3

| Grey level | $T_1 = OR(t_{2,1}, t_{3,1})$ | $T_2 = OR(t_{2,2}, t_{3,2})$ | $T_3 = OR(t_{2,3}, t_{3,3})$ | $\tilde{P} = \overline{T_1 \otimes T_2 \otimes T_3}$ |
|---|---|---|---|---|
| 0 | 100100 | 010010 | 001001 | 000000 |
| 1 | 100011 | 010101 | 001110 | 000111 |
| 2 | 011011 | 101101 | 110110 | 111111 |

From the above table, we can compute the contrast $\alpha_0 = \frac{3-0}{6} = \frac{1}{2}$, $\alpha_1 = \frac{6-3}{6} = \frac{1}{2}$. We get $\alpha_0 = \alpha_1$, namely the contrasts between every grey level are the same, this makes the reconstructed secret image has higher visual quality.

# Appendix D. Analysis and experimental results of the section 5

We first briefly describe Hu and Tzeng's PBRVCS in the appendix, and then give an example to show that Hu and Tzeng's PBRVCS cannot be directly extended to grayscale RVCS.

## 1. Hu and Tzeng's' RVCS for binary image

We show construction procedures of Hu and Tzeng' scheme in Table D-1 as follows. Employing basis matrices $B_0$ and $B_1$ of an optimal Naor and Shamir's (*k*, *k*)-nRVCS, which is perfect black VCS.

Hu and Tzeng's gave a construction novel basis matrices and an auxiliary matrix to create a (k, n)-RVCS for binary image.

$L_0$, $L_1$ and $A_0$ are two basis matrices and an auxiliary matrix, respectively in a Hu and Tzeng's $(k,n)-BVCS$.

$L_0 = E_1^0 \circ \cdots \circ E_v^0, L_1 = E_1^1 \circ \cdots \circ E_v^1$, where $v = \binom{n}{k}$ .

$E_j^0$ ($E_j^1$), the $j_1$-th, … $j_k$-th rows is the 1st , ..., *k*th rows of $B_0$ ($B_1$),the elements of other rows of $E_j^0$ （$E_j^1$） are all 1's . $(j_1, \cdots, j_k) \subseteq (1, \cdots, k)$.

$E_j^0$ （$E_j^1$）, so the two matrices $L_0$ and $L_1$ have $n \times (v \cdot 2^{k-1})$ size.

**Auxiliary matrix** $A_0$:The construction of $A_0$ is similar to $F_p$ and $L_1, A_0 = F_1 \| \cdots \| F_v$, the elements in $j_1$-*th* , … $j_k$-*th* rows is the 1st , ..., k*th* rows of $B_0$ ($B_1$) are all 0's , the other rows of $F_p$ are all 1's. In other words, $A_0$ is the same matrix as $L_0$ ($L_1$) except that we replace all the elements of the corresponding that of $(k,k)-nRVCS$ with all 0's

Let $C_p^0$ and $C_p^1$ be the collection of basis Boolean matrices $E_p^0$ and $E_p^1$, where $1 \le p \le v$. Let $C_p^A$ be the collection of Boolean matrix $F_p$ define as above.

The dealer encodes each transparency $t_i$ as $v$ sub-transparencies $t_{i,p}$ and each sub-block consists of one secret image. For $1 \le p \le v$, each white (black pixel) on sub-block $t_{i,p}$ is encoded using $n \times 2^{k-1}$ matrices $E_p^0$ ($E_p^1$ resp.). Table D-1 shows the construction of Hu and Tzeng's scheme.

Table D-1 Distribution phase and reconstruction phase of Hu and Tzeng's scheme

| **Distribution phase** | **Reconstruction phase** |
|---|---|
| To share a white(black, resp.) pixel, the dealer,<br>**Step1:** randomly chooses a matrix $S_p^0 = [s_{ij}]$ in $C_p^0$ ( $S_p^1$ in $C_p^1$ resp.) , and a matrix $A_p^0 = [a_{i,j}]$ in $C_p^A$.<br>**Step 2:** For each participant $i$, put a white (black, resp.) pixel on the sub-block $t_{i,j}$ if $s_{i,j} = 0$, ($s_{i,j} = 1$ resp.).<br>**Step 3:** For each participant $i$, put a white (black, resp.) pixel on the sub-block $A_{i,j}$ if $a_{i,j} = 0$, ($a_{i,j} = 1$ resp.). | Any $k$ participants in $Q$ reconstruct the secret image by computing:<br>**Step 1:**XORing all the shares $t_j$ and stacking all the shares $A_j$ for $j = 1, \ldots, k_p$ and obtain $T$ and $A$ respectively.<br>**Step 2:** $U = (T \oplus A) \otimes A$ , $\tilde{P} = U$ is the reconstructed secret image. |

***Example D-1*:** A Hu and Tzeng's (2, 3) scheme.

The Basic matrix $B_0$ and $B_1$ in a (2, 2)-VCS are

$$B_0 = \begin{bmatrix} 10 \\ 10 \end{bmatrix}, B_1 = \begin{bmatrix} 10 \\ 01 \end{bmatrix}.$$

$$E_1^0 = \begin{bmatrix} 10 \\ 10 \\ 11 \end{bmatrix}, E_2^0 = \begin{bmatrix} 11 \\ 10 \\ 10 \end{bmatrix}, E_3^0 = \begin{bmatrix} 10 \\ 11 \\ 10 \end{bmatrix}.$$

$$E_1^1 = \begin{bmatrix} 10 \\ 01 \\ 11 \end{bmatrix}, E_2^1 = \begin{bmatrix} 11 \\ 10 \\ 01 \end{bmatrix}, E_3^1 = \begin{bmatrix} 10 \\ 11 \\ 01 \end{bmatrix}.$$

$$F_1 = \begin{bmatrix} 00 \\ 00 \\ 11 \end{bmatrix}, F_2 = \begin{bmatrix} 11 \\ 00 \\ 00 \end{bmatrix}, F_3 = \begin{bmatrix} 00 \\ 11 \\ 00 \end{bmatrix}.$$

$$L_0 = \begin{bmatrix} 101110 \\ 101011 \\ 111010 \end{bmatrix}, L_1 = \begin{bmatrix} 101110 \\ 011011 \\ 110101 \end{bmatrix}, A_0 = \begin{bmatrix} 001100 \\ 000011 \\ 110000 \end{bmatrix}.$$

Table D-2 The distribution phase of Hu and Tzeng's (2, 3) scheme

| Pixel | 1st run | 2nd run |
|---|---|---|
| Black | $t_1$ =101110 | $A_1$ =001100 |
| | $t_2$ =011011 | $A_2$ = 000011 |
| | $t_3$ =110101 | $A_3$ =110000 |
| White | $t_1$ =101110 | $A_1$ = 001100 |
| | $t_2$ =101011 | $A_2$ = 000011 |
| | $t_3$ =111010 | $A_3$ = 110000 |

Table D-3 The reconstruction phase of Hu and Tzeng's (2, 3) scheme

| Pixel | 1st run | 2nd run | |
|---|---|---|---|
| | Participant 1 + participant 2 | | |
| Pixel | 1st run $T = t_1 \otimes t_2$ | 2nd run $A = A_1 \otimes A_2$ | $\tilde{P}$ |
| Black | 110101 | 001111 | 110000 |
| White | 000101 | 001111 | 000000 |
| | Participant 1 + participant 3 | | |
| Pixel | 1st run $T = t_1 \otimes t_3$ | 2nd run $A = A_1 \otimes A_3$ | $\tilde{P}$ |
| Black | 011011 | 111100 | 000011 |
| White | 010100 | 111100 | 000000 |
| | Participant 2 + participant 3 | | |
| Pixel | 1st run $T = t_2 \otimes t_3$ | 2nd run $A = A_2 \otimes A_3$ | $\tilde{P}$ |
| Black | 101110 | 110011 | 001100 |
| White | 010001 | 110011 | 000000 |

The contrasts between every grey level are as follows:
Participant 1 + participant 2: $\alpha = (2-0)/2 = 1$,
Participant 1 + participant 3: $\alpha = (2-0)/2 = 1$
Participant 2 + participant 3: $\alpha = (2-0)/2 = 1$.

From the reconstruction phase above, any two participants of three participants can correctly recovery original pixel.

**The decoding complexity for Hu and Tzeng's scheme.**

The reconstruction phase of the $(k,n)$-PBRVCS, operations of stacking any $k$ shares equal $4k$ OR operations. Then $4k-1$ NOT operations are required to finish 2 runs. Each participant holds 2 shares. The size of shares becomes $2^k\binom{n}{k}$ times larger than that of the original secret image. The size of the reconstructed image is the same as that of the original image.

## 2. *The analysis of directly extending RVCS above to RGVCS*

Now we give an example to show that Hu and Tzeng's PBRVCS cannot be directly extended to grayscale RVCS.

***Example D-2 (continuation of D-1) :*** directly extend Hu et al.'s binary (2, 3)-RVCS to GVCS. The basis matrices and Auxiliary matrix for three grey levels (2, 3)-GVCS are:

$$L^0 = \begin{bmatrix} \overbrace{1010}^{\{1,2\}} \\ 1010 \\ 1111 \end{bmatrix} \circ \begin{bmatrix} \overbrace{1010}^{\{1,3\}} \\ 1111 \\ 1010 \end{bmatrix} \circ \begin{bmatrix} \overbrace{1111}^{\{2,3\}} \\ 1010 \\ 1010 \end{bmatrix}, L^1 = \begin{bmatrix} \overbrace{1010}^{\{1,2\}} \\ 1001 \\ 1111 \end{bmatrix} \circ \begin{bmatrix} \overbrace{1010}^{\{1,3\}} \\ 1111 \\ 1001 \end{bmatrix} \circ \begin{bmatrix} \overbrace{1111}^{\{2,3\}} \\ 1010 \\ 1001 \end{bmatrix},$$

$$L^2 = \begin{bmatrix} \overbrace{1010}^{\{1,2\}} \\ 0101 \\ 1111 \end{bmatrix} \circ \begin{bmatrix} \overbrace{1010}^{\{1,3\}} \\ 1111 \\ 0101 \end{bmatrix} \circ \begin{bmatrix} \overbrace{1111}^{\{2,3\}} \\ 1010 \\ 0101 \end{bmatrix}, GA = \begin{bmatrix} \overbrace{0000}^{\{1,2\}} \\ 0000 \\ 1111 \end{bmatrix} \circ \begin{bmatrix} \overbrace{0000}^{\{1,3\}} \\ 1111 \\ 0000 \end{bmatrix} \circ \begin{bmatrix} \overbrace{1111}^{\{2,3\}} \\ 0000 \\ 0000 \end{bmatrix}.$$

For every pixel, the dealer randomly chooses matrices, which are gotten by doing totally random column permutation to the basis matrices, to distribute the pixel in each share.

Table D-4 The distribution phase of directly extending Hu and Tzeng's (2, 3) scheme

| Grey levels | Chosen matrices | 1st run | 2nd run |
|---|---|---|---|
| 0 | $\begin{bmatrix} 100101101111 \\ 110101111010 \\ 101111101010 \end{bmatrix}$ | $t_1$=100101101111<br>$t_2$=110101111010<br>$t_3$=101111101010 | $A_1$=000000001111<br>$A_2$= 000011110000<br>$A_3$=111100000000 |
| 1 | $\begin{bmatrix} 100101101111 \\ 110011111010 \\ 101111011001 \end{bmatrix}$ | $t_1$=100101101111<br>$t_2$=110011111010<br>$t_3$=101111011001 | $A_1$=000000001111<br>$A_2$= 000011110000<br>$A_3$=111100000000 |
| 2 | $\begin{bmatrix} 100101101111 \\ 011011111010 \\ 111110010101 \end{bmatrix}$ | $t_1$=100101101111<br>$t_2$= 011011111010<br>$t_3$=111110010101 | $A_1$=000000001111<br>$A_2$= 000011110000<br>$A_3$=111100000000 |

Table D-5 The reconstruction phase of directly extending Hu and Tzeng's (2, 3) scheme

| Grey levels | Participant 1 + participant 2 | | $\tilde{P}$ |
|---|---|---|---|
| | 1st run<br>$T = t_1 \otimes t_2$ | 2nd run<br>$A = A_1 \otimes A_2$ | |
| 0 | 010000010101 | 000011111111 | 010000000000 |
| 1 | 010110010101 | 000011111111 | 010100000000 |
| 2 | 111110010101 | 000011111111 | 111100000000 |

| Participant 1 + participant 3 | | | |
|---|---|---|---|
| Grey levels | 1st run $T = t_1 \otimes t_3$ | 2nd run $A = A_1 \otimes A_3$ | $\tilde{P}$ |
| 0 | (001010000101) | (111100001111) | (000010000000) |
| 1 | (001010110110) | (111100001111) | (000010110000) |
| 2 | (011011111010) | (111100001111) | (000011110000) |

| Participant 2 + participant 3 | | | |
|---|---|---|---|
| Grey levels | 1st run $T = t_2 \otimes t_3$ | 2nd run $A = A_2 \otimes A_3$ | $\tilde{P}$ |
| 0 | (011010010000) | (111111110000) | (000000000000) |
| 1 | (011100100011) | (111111110000) | (000000000011) |
| 2 | (100101101111) | (111111110000) | (000000001111) |

The contrasts between every grey level are as follows:

Participant 1 + participant 2: $\alpha^{(1,0)} = 1/4$, $\alpha^{(2,1)} = (4-2)/4 = 1/2$,

Participant 1 + participant 3: $\alpha^{(1,0)} = (3-1)/4 = 1/2$, $\alpha^{(2,1)} = (4-3)/4 = 1/4$,

Participant 2 + participant 3: $\alpha^{(1,0)} = (2-0)/4 = 1/2$, $\alpha^{(2,1)} = (4-2)/4 = 1/2$.

In the reconstruction process above, we get the contrasts between every grey level are inconsistent, which will affect the quality of the reconstructed secret image.

***Example D- 3 (continuation of Example 5.1 in section 5):***

Table D-6 Distribution phase for the dealer

| Grey levels | Basis matrices | 1st run | 2nd run |
|---|---|---|---|
| 1 | $\begin{bmatrix} 1&0&1&0&1&0&1&0&1&1&1&1 \\ 1&0&1&0&1&1&1&1&1&0&1&0 \\ 1&1&1&1&1&0&1&0&1&0&1&0 \end{bmatrix}$ | $t_1$ =1010\|1010\|1111<br>$t_2$ =1010\|1111\|1010<br>$t_3$ =1111\|1010\|1010 | $A_1$ =0000\|0000\|1111<br>$A_2$ = 0000\|1111\|0000<br>$A_3$ =1111\|0000\|0000 |
| 2 | $\begin{bmatrix} 1&0&1&0&1&0&1&0&1&1&1&1 \\ 1&0&0&1&1&1&1&1&1&0&1&0 \\ 1&1&1&1&1&0&0&1&1&0&0&1 \end{bmatrix}$ | $t_1$ =1010\|1010\|1111<br>$t_2$ =1111\|1001\|1001<br>$t_3$ =1010\|1010\|1111 | $A_1$ =0000\|0000\|1111<br>$A_2$ = 0000\|1111\|0000<br>$A_3$ =1111\|0000\|0000 |
| 3 | $\begin{bmatrix} 1&0&1&0&1&0&1&0&1&1&1&1 \\ 0&1&0&1&1&1&1&1&1&0&1&0 \\ 1&1&1&1&0&1&0&1&0&1&0&1 \end{bmatrix}$ | $t_1$ =1010\|1111\|1010<br>$t_2$ =0101\|1010\|1111<br>$t_3$ =1111\|0101\|0101 | $A_1$ =0000\|0000\|1111<br>$A_2$ = 0000\|1111\|0000<br>$A_3$ =1111\|0000\|0000 |

Table D-7 Reconstruction phase

| Participant 1 + Participant 2 | | | |
|---|---|---|---|
| Grey levels | 1st run | 2nd run | $\tilde{P}$ |
| | $T = t_1 \otimes t_2$ | $A = A_1 \otimes A_2$ | |
| 1 | 0000\|0101\|0101 | 0000\|1111\|1111 | 0000\|0000\|0000 |
| 2 | 0011\|0101\|0101 | 0000\|1111\|1111 | 0011\|0000\|0000 |
| 3 | 1111\|0101\|0101 | 0000\|1111\|1111 | 1111\|0000\|0000 |
| | Participant 1 + Participant 3 | | |
| Grey levels | 1st run<br>$T = t_1 \otimes t_3$ | 2nd run<br>$A = A_1 \otimes A_3$ | $\tilde{P}$ |
| 1 | 0101\|0000\|0101 | 1111\|0000\|1111 | 0000\|0000\|0000 |
| 2 | 0101\|0011\|0110 | 1111\|0000\|1111 | 0000\|0011\|0000 |
| 3 | 0101\|1111\|1010 | 1111\|0000\|1111 | 0000\|1111\|0000 |
| | Participant 2 + Participant 3 | | |
| Grey levels | 1st run<br>$T = t_2 \otimes t_3$ | 2nd run<br>$A = A_2 \otimes A_3$ | $\tilde{P}$ |
| 1 | 0101\|0101\|0000 | 1111\|1111\|0000 | 0000\|0000\|0000 |
| 2 | 0110\|0110\|0011 | 1111\|1111\|0000 | 0000\|0000\|0011 |
| 3 | 0101\|1010\|1111 | 1111\|1111\|0000 | 0000\|0000\|1111 |

The contrasts between every grey level are as follows:

Participant 1 + participant 2: $\alpha^{(1,0)} = 1/2$, $\alpha^{(2,1)} = 1/2$,

Participant 1 + participant 3: $\alpha^{(1,0)} = 1/2$, $\alpha^{(2,1)} = 1/2$,

Participant 2 + participant 3: $\alpha^{(1,0)} = 1/2$, $\alpha^{(2,1)} = 1/2$.

From the reconstruction process above, the contrasts between every grey level are the same, this means the reconstructed secret image has higher visual quality.